\documentclass[useAMS,usenatbib]{mn2e}
\usepackage{graphicx}
\def\HST{HST}
\def\Spitzer{{\it Spitzer}}
\def\spitzer{{\it Spitzer}}
\def\hst{HST}
\def\sst{{\it Spitzer}}

\title[Dust in the atmosphere of HD 189733b]{The prevalence of dust on the exoplanet HD 189733b from Hubble and Spitzer observations}
\author[F. Pont et al.]{F. Pont$^{1}$\thanks{E-mail: fpont@astro.ex.ac.uk}, D. K. Sing$^{1}$, N. P. Gibson$^{2,3}$,  S. Aigrain$^{2}$,  G. Henry$^{4}$, N. Husnoo$^{1}$\smallskip
\\
$^{1}$School of Physics, University of Exeter, Stocker Road, Exeter, EX4 4QL, UK\\
$^{2}$Department of Physics, University of Oxford, Denys Wilkinson Building, Keble Road, Oxford OX1 3RH, UK\\
$^{3}$European Southern Observatory, Karl-Schwarzschild-Str. 2, 85748 Garching bei M\"unchen, Germany\\
$^{4}$Tennessee State University, CEIS, 3500 John A. Merritt Blvd., P.O. Box 9501, Nashville, TN 37209, USA}
\begin{document}

\date{Accepted  Received ; in original form}

\pagerange{\pageref{firstpage}--\pageref{lastpage}} \pubyear{2012}

\maketitle

\label{firstpage}

\begin{abstract}

The hot Jupiter HD 189733 b is the most extensively observed exoplanet. Its atmosphere has been detected and characterised in transmission and eclipse spectroscopy, and its phase curve measured at several wavelengths. This paper brings together the results of our campaign to obtain the complete transmission spectrum of the atmosphere of this planet from UV to infrared with the Hubble Space Telescope, using the STIS, ACS and WFC3 instruments. We provide a new tabulation of the transmission spectrum across the entire visible and infrared range. The radius ratio in each wavelength band was re-derived, where necessary, to ensure a consistent treatment of the bulk transit parameters and stellar limb-darkening. Special care was taken to correct for, and derive realistic estimates of the uncertainties due to, both occulted and unocculted star spots.

The combined spectrum is very different from the predictions of cloud-free models for hot Jupiters: it is dominated by Rayleigh scattering over the whole visible and near infrared range, the only detected features being narrow sodium and potassium lines. We interpret this as the signature of a haze of condensate grains extending over at least five scale heights. We show that a dust-dominated atmosphere could also explain several puzzling features of the emission spectrum and phase curves, including the large amplitude of the phase curve at 3.6 \micron, the small hot-spot longitude shift and the hot mid-infrared emission spectrum. We discuss possible compositions and derive some first-order estimates for the properties of the putative condensate haze/clouds. We finish by speculating that the dichotomy between the two observationally defined classes of hot Jupiter atmospheres, of which HD 189733 b and HD 209458 b are the prototypes, might not be whether they possess a temperature inversion, but whether they are clear or dusty. We also consider the possibility of a continuum of cloud properties between hot Jupiters, young Jupiters and L-type brown dwarfs.

\end{abstract}


\section{Introduction}




Transiting exoplanets are giving us our first glimpse into planetary atmospheres beyond the Solar System. Observational biases make short-orbit gas giants (hot Jupiters) most accessible to observations \citep[see ][for a textbook-level introduction]{has10}. The planet HD 189733b \citep{bou05,bak06,ago10} occupies a place of choice among the observable targets, because of its bright host star and large planet-to-star radius ratio.  HD 189733b has been observed abundantly with both the Hubble and Spitzer space telescopes in order to characterise its atmosphere. It epitomises both the promises and challenges of these observations.

\subsection{The hot Jupiter HD 189733b}
HD 189733b is a 1.15 M$_{\rm Jup}$ gas giant, orbiting its K-dwarf host star in 2.2 days. Due to the relatively small size of the star, a 0.77 $R_\odot$ K dwarf,  the transit signal is deep (2.4\%). Three types of observations can be used to characterise the atmosphere of the planet: the transmission spectrum along the limb during transits, the day-side emission spectrum during secondary eclipses, and the day-night temperature contrast with the orbital phase curve. One atmospheric scale height projects to about 1$\times 10^{-4}$ in transmission spectroscopy. The secondary eclipse is 3$\times 10^{-3}$ deep at 8~\micron, and the phase curve amplitude is $1 \times 10^{-3}$ at 4 \micron.
Data on the atmosphere of HD 189733b using these three methods have been gathered using the Hubble Space Telescope (``\hst ''), the Spitzer Space Telescope (``\sst '') and large ground-based telescopes.

\subsection{Expectations from atmosphere models}
The irradiation temperature (equilibrium temperature at the sub-stellar point) of HD 189733b is 1700~K, making the zero-albedo equilibrium temperature 1200 to 1400~K depending on the efficiency of heat transfer from the day side to the night side. At these temperatures, the hot jupiter atmosphere models of \citet{for08} predict a clear, dark atmosphere, dominated by absorption by neutral sodium and potassium in the visible, and molecular bands of water, ammonia, carbon monoxide, carbon dioxide and methane in the infrared. In the blue and near UV, scattering by hydrogen molecules is expected to become the dominant source of opacity. In chemical-equilibrium models, the temperature in the atmosphere is too cold to sustain titanium oxide and vanadium oxide vapours, that can provide high opacities in the visible for hotter atmospheres. Vigorous vertical mixing could bring these species up from the hotter depths \citep[see e.g.][]{spi09}. 

The dominant source of heat in the atmosphere is the stellar irradiation, which far exceeds the internal heat leakage from the interior of the planet. As a result of the injection of heat from above, the atmosphere is expected to be vertically stable, with radiative transfer being the dominant mechanism of vertical energy exchange \citep{sho02}. This would tend to inhibit the formation of dust clouds by condensation. However, the atmospheric temperature of the planet is close to the condensation temperature of several abundant components, including silicates and iron. The possible presence of dust was considered early on \citep{sea98}. According to models, condensates would make the spectral features weaker, or mask some of them, depending on the height of the cloud deck \citep{mar99,sea00,sud03}. \citet{for05} studied the effect of condensates on the transmission spectrum specifically. That paper proposes that high-altitude condensates may be ubiquitous in hot Jupiter atmospheres and concludes with the following statement: ``We assert that transmission spectroscopy will continue to yield abundances of expected chemical species far below those predicted for a ÔclearÕ atmosphere''

Smaller grains or partly transparent hazes can also imprint some features on the spectrum, such as Rayleigh scattering in the blue or silicate absorption features in the mid-infrared.

Photochemistry may also play a role in producing condensates, for instance sulphur or carbon compounds, that could affect the spectral signatures of the planetary atmosphere \citep[e.g.][]{zah09}.


\subsection{Observation of the atmosphere of HD 189733b} 
The transit spectrum of HD 189733b has been measured from 300 nm to 1 \micron\ with the STIS and ACS instruments aboard \hst, from 1 to 3 \micron\ with \hst's WFC3 and NICMOS \citep{swa08,pon08,sin09,des09,sin11,gib11}, and in five infrared passbands between 3.6 and 24 \micron\ with IRAC and MIPS on \spitzer\ \citep{ehr07,des09,knu07,knu09,ago10}. The emission spectrum has been measured in the five IRAC and MIPS passbands and from 8 to 13 \micron\ with \spitzer's IRS \citep{ago10,gri08,des09,des11}. The phase curve along the orbit has been measured in the 3.6, 4.5 and 8 \micron\ channels of IRAC, and around 24 \micron\ with MIPS \citep{knu07b,knu09,knu12}.
The signature of Rayleigh scattering \citep{pon08,sin11} and the sodium doublet \citep{red08,hui12} have been detected in the UV and visible.
Claims of detection of molecular features in the infrared \citep{swa08} have not been confirmed by subsequent measurements and analyses \citep{sin09,des09,des11,gib11,gib12b}. 

Inferring atmospheric properties  from the present observations is a challenge. The signals remain small even by the standard of space observations, and the uncertainties are generally dominated by complex instrumental effects. It has not been uncommon for new observations, and even new analysis of existing observations, to contradict earlier results. In the infrared, the wavelength coverage is sparse (imposed by the \sst\ passbands in space and the water absorption gaps from the ground). With a handful of passbands for as many molecules, interpretation often admits several possibilities.


In the emission spectrum, the planetary flux data in \Spitzer\ passbands were compatible with a temperature profile decreasing with height and the presence of water and other molecules. However, new phase curves at 3.6 and 4.5~\micron\ and a re-analysis of the previous \Spitzer\ data by \citet{knu12} have also overhauled previous results on the emission spectrum. The eclipse depth at 3.6 \micron\ from \citet{cha08}, crucial to constrain the temperature profile and water abundance, turn out to be incorrect by more than 3 sigma, due to the complex \sst\ instrumental systematics (the ``ramp effect'' and ``pixel phase effect'' affecting high-accuracy flux measurements). The amplitude of the phase curves and depth of the eclipses at 3.6 and 4.5 \micron\ are no longer amenable to a simple interpretation. \citet{knu12} mention non-equilibrium chemistry and slowed day-night recirculation as possible explanations of the anomalies in the data.

Overall, the current interpretation of the ensemble data is of an atmosphere corresponding to dust-free models with molecular absorption and without a stratospheric temperature inversion. A thin layer of haze at high altitude produces the Rayleigh scattering signature in the visible in the transmission spectrum, but because the grains are small, it does not affect the emission spectrum, the transmission spectrum in the infrared or the global energy budget of the atmosphere.

\subsection{HST programme GO-11740 and present paper}

Our HST programme with STIS and WFC3 (GO-11740) was designed to put this interpretation to the test. This programme was aimed at filling the wavelength gaps in transmission spectroscopy, providing a continuous coverage from the UV to near infrared to address the discrepancies between the different datasets and lift some ambiguities in interpretations.  It provides the observational basis of the present paper. The explanation in terms of a thin layer of haze at high altitude requires the transmission spectrum to evolve sharply from a featureless slope to deep molecular features in the 1-2 \micron\ interval between the ACS and NICMOS measurements.  The STIS measurements have shown the  featureless slope to extend beyond the range probed initially by ACS, suggesting that Rayleigh scattering dominates the transmission spectrum over at least five atmospheric scale heights. The finding that Rayleigh scattering dominates the transmission spectrum continuously from the UV to the near infrared opens the possibility that the optically thick haze layer is more than a marginal phenomenon affecting only the very low-pressure parts of the atmosphere, and may affect the radiative transfer and spectrum throughout the atmosphere. 

In this paper, we present the global results of the HST GO-11740 programme to obtain a complete transmission spectrum from UV to near-IR with \hst. Results for individual instruments were presented in \citet{gib12} and \citet{sin11}. We concentrate here on combining the different wavelength ranges into a single coherent scale. This is rendered difficult primarily because HD 189733 presents variability at the 1-2\% level in the visible due to star spots, so that measuring the transmission spectrum of the planet implies first disentangling the effect of the star spots, which has a similar amplitude.  We then consider the whole ensemble of data on the atmosphere of HD 189733b, including the key update by \citet{knu12} on secondary eclipses and phase curves. We explore the possibility that the dust detected in the transmission spectrum has a large impact on the atmosphere as a whole. 

Section \ref{go} presents the collection and modelling of the transit measurements of the GO-11740 programme and other programmes pertaining to the  transmission spectrum. Section \ref{starspots} tackles the crucial but somewhat technical issue of accounting for star spots on the parent  star. Section \ref {secresults} presents the global UV-to-infrared transmission spectrum obtained by combining the datasets in a consistent manner.  Section \ref{spec} discusses the influence of haze and clouds in the atmosphere data as a whole, including emission and phase curve information. In Section \ref{toy}, we explore the possibility that of the atmosphere of HD 189733b is dominated by condensates, and examine the observable consequences.


\section{Atmosphere transmission spectrum: collecting and modelling transit observations}

The transmission spectrum of the atmosphere of a transiting planet can be measured by monitoring the transit in spectroscopy. In principle the method is simple: since the depth of the transit is proportional to the square of the planet-to-star radius ratio, measuring the depth of the transit at different wavelengths records the opacity of the atmosphere to grazing star light at those wavelengths, yielding a transmission spectrum. 

As outlined in the Introduction, the transit depth of the HD 189733 system has been measured precisely over many wavelengths from 0.3 to 24 \micron. On paper, the  potential of these measurements is impressive. The expected amplitude of the atmospheric features is about $5 \cdot 10^{-4}$ as a fraction of the stellar flux, and the photon-shot noise integrated over the whole transit is smaller than $1 \cdot 10^{-4}$ at all wavelengths up to 8 \micron. Such measurements are able to constrain the atmospheric structure and composition in significant detail. 

However, in practice, transit spectroscopy of HD~189733b has had to contend with two obstacles. The first is the ubiquity of complex instrumental effects at and above the $10^{-4}$ level in flux measurements, particularly with infrared detectors. The second obstacle is the fact that HD 189733 is an active, spotted star, which introduces an additional signal in the relation between transit depth and planet size.

These issues are discussed in \citet{pon08}, \citet{des09}, \citet{sin11} and \citet{gib11}. They prevent an easy comparison of the transit radius in different wavelength ranges, so that producing a global estimate of the UV-to-IR transmission spectrum requires a detailed combined analysis of all data sets. This analysis makes up the rest of the present Section.

In contemplating the possibility of measuring a global transmission spectrum for HD 189733b from measurements widely separated in time, we make the assumption that intrinsic variations in the atmosphere of the planet over time are smaller than the present observational uncertainties.  This assumption is discussed in Paragraph \ref{weather}.

\label{go}

\subsection{Transmission spectroscopy datasets}

Table \ref{listobs} gives the journal of the space-based observations of the transit of HD 189733b considered in this paper. The primary focus of this study is addressed by our spectroscopic measurements with the STIS, ACS and WFC3 instruments on \HST, to provide a continuous transmission spectrum from the near UV (300 nm) to the near IR (1.2~\micron). This covers the interval between high-altitude atomic lines in the UV and molecular lines near the photosphere in the infrared, encompassing the part of the atmosphere involved in the bulk of the energy exchange between the stellar irradiation and the planetary atmosphere. We also include spectroscopic and photometric measurements with the NICMOS camera on \HST, as well as broadband photometric measurements with the IRAC and MIPS instruments on \Spitzer. 

The individual datasets, presented separately in the publications mentioned in the previous section, are briefly summarised below.  

\subsubsection{STIS (HST)}

Five transits of HD 189733 were observed with STIS on the refurbished HST in two contexts, at low resolution from the present programme, and at medium resolution around the sodium doublet (GO-11111). The results of the first observations are published in \citet{sin11} and the second in \citet{hui12}. The broad-band data showed a featureless spectrum over the 300-600 nm range, with absorption rising bluewards. The medium-resolution data resolved both components of the sodium doublet, showing strong narrow cores and an absence of broad wings for the sodium feature. 

\subsubsection{ACS (HST)}

Three transits were observed\ with the ACS camera on HST, using Grism  G800L (0.6-1 \micron) on May 22, May 26 and July 14, 2006. The results were published in \citet{pon07} for the time series and \citet{pon08} for the transmission spectrum.

One important issue for the present study is the measurement of
potassium absorption. A very large line with extended wings is
expected in a clear atmosphere from theoretical models, but was not {learly
seen in the broadband data analysis of \citet{pon08}. The ACS
measurements do not have the resolution to detect the narrow core of
the potassium line, and is only sensitive to its total equivalent
width over a larger wavelength band.   Here we present a new pixel-by-pixel
reanalysis of the ACS data, with the intent of placing more stringent constraints on
the presence of potassium.

We find excess absorption located close to the expected position of the core of the potassium doublet, but the resolution is too low and the noise on pixel-by-pixel decorrelation of systematics too high to give a measurement of the width of the detection. We therefore fit the amplitude of a potassium line model to the pixel-by-pixel data, with the  model used in \citet{hui12} to fit the potassium line and surrounding Rayleigh scattering slope, which is based upon the analytic expression from \citet{lec08}.  As with the sodium feature, for the potassium cross-section we neglected pressure broadening due to the observed lack of broad line wings.  We fixed the abundance to an arbitrary level, which is unconstrained as the haze has an unknown composition, and fit for the baseline altitude of a 900~\AA\ spectral region surrounding the sodium doublet, as well as the model temperature, assumed to be isothermal.  A 900~\AA\ width was chosen such that the baseline altitude was well measured, while also ensuring that the best-fit model temperature was not sensitive to the broadband Rayleigh slope. The result is shown on Fig. \ref{ACSK}. The potassium feature in the ACS data is 2.5-$\sigma$ significant. The model temperature scales to fit the amplitude of the potassium feature, with a best-fit value of 1800$\pm$720 K.
  
  Thus, while the lack of high-resolution data around the potassium line
prevents very specific constraints, the best-fit solution is
compatible with the presence of the line to the same level as the
sodium line or higher, i.e. with a possible line core but no broad wings.  The possible presence of potassium in the ACS data is in contrast to the findings of \citet{jen11} who did not detect potassium using high-resolution ground-based spectroscopy.   However, the different resolutions, wavelength range, and planetary altitudes probed combined with the low signal-to-noise of both studies makes a direct comparison difficult.

\begin{figure}
{ \centering                            
\includegraphics[width=0.5\textwidth,trim=-0.2in -0.2in 0.2in -0.1in,clip=true]{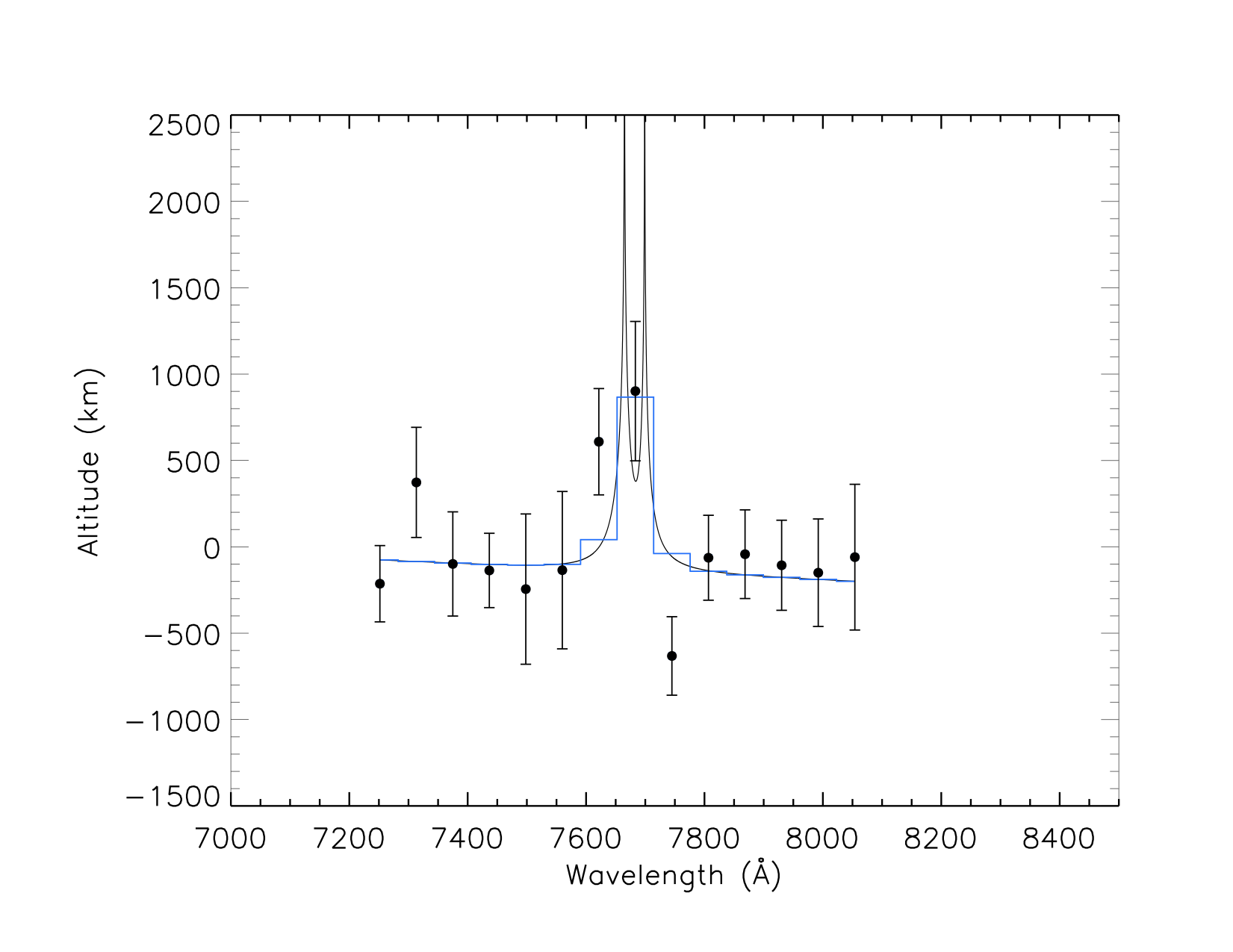}}
\caption[]{ACS transmission spectrum of HD~189733b in a 900 \AA\ region surrounding the potassium doublet.  Also plotted is a best-fit model (both unbinned and binned to the data) which includes Rayleigh scattering and sodium absorption.}
\label{ACSK}
\end{figure}

\subsubsection{ WFC3 (HST)}

Two transits were observed with the WFC3 camera on the refurbished HST as part of our programme GO-11740, covering the wavelength interval 1.1--1.7 \micron. The results are presented in \citet{gib12}. Due to saturation and non-linearity affecting the brightest (central) pixels of the spectrum, light curves were extracted from the blue and red ends of the spectra only, corresponding to wavelength ranges of 1.099--1.168 and 1.521--1.693 \micron\ for the first visit, and 1.082--1.128 and 1.514--1.671 \micron\ for the second. To account for instrumental systematics, the light curves were fitted using a Gaussian-process model whilst simultaneously fitting for the transit parameters.


\subsubsection{NICMOS (HST)}

Transit spectroscopy data for HD189733b has been gathered by NICMOS in two modes: spectroscopic and photometric. The first is presented in \citet{swa08}, the second in  \citet{sin09}. The two datasets reach incompatible conclusions. \citet{gib11}, and \citet{gib12b} reanalyse the spectroscopic data, and discuss this issue. The reader is referred to these papers for details.  We adopt the method of \citet{gib12b}  to calculate the uncertainties of the spectroscopic data. These use a Gaussian process to model the instrumental systematics, which avoids the restrictive assumption of linear basis models and performs the inference in a Bayesian way, therefore mitigating against over-interpreting systematics.

\subsubsection{IRAC (\sst)}

We use \citet{des09}, \citet{des11} and \citet{knu12} for measurements in the 3.6-\micron, 4.5-\micron\ and 5.8 \micron\ channels of IRAC on Spitzer, and \citet{ago10} for the 8-micron channel.


\subsection{Self-consistent transit modelling}

Three issues must be addressed in  order to place all the observations of the radius of HD 189733b on a common scale and build a transmission spectrum over a wide wavelength range: the orbital elements, the stellar limb darkening, and the effect of stellar variability and star spots. 


\subsubsection{Orbital parameters}

These measurements of radius ratio used to construct the transmission spectrum are determined as part of a model transit light curve fit, that also includes other parameters of the system, namely the orbital ephemerides, the stellar density and the orbital inclination. Any offset in these quantities will result in a difference in the inferred radius ratio. This difference will be slightly wavelength-dependent, first, because of the corresponding changes in the effect of stellar limb darkening, and second, because different parts of the transit were measured with different instruments, so that the covariance between the transit depth and other orbital parameters will vary. Because different values for the system parameters have been used in separate studies, the resulting planet radius values cannot be compared directly.

Fortunately, in the case of HD189733b, the orbital parameters are
known with extreme precision. This is therefore not a significant
source of uncertainty in the final results. The best measurements of
the orbital parameters come from the monitoring of 14 transits and
eclipses at 8-\micron\ with Spitzer \citep[][see Table
\ref{table1}]{ago10}. We re-calculated the radius ratio for all data
sets using these ephemerides.  In the Spitzer bands where the effects of limb-darkening
 are small, we corrected the values calculated with other parameters using the first-order expressions of the relation between orbital parameters and transit shape given in \citet{car08}. Otherwise we re-fitted the original data with the reference system parameters.

\begin{table}
\begin{tabular}{lc} \hline
Transit central time & 2454279.436714 [BJD]\\
Orbital period & 2.21857567 days\\
Impact parameter & 0.6631\\
Tangential velocity & 25.125 R$_* \,{\rm days}^{-1}$\\ \hline
\end{tabular}
\caption{Orbital parameters adopted for the re-analysis in this paper, from \citet{ago10}: timing of transit, orbital period, impact parameter, tangential velocity during the transit. }
\label{table1}
\end{table}


\begin{table}
\begin{tabular}{llrc}
\hline
Instrument & Grism or & Number of  & Wavelength \\
  &  filter & transits used &  range [nm] \\
 \hline
STIS & G430L & 2 & 290 -- 570  \\
STIS & G750M & 3 & 580 -- 637  \\
ACS &G800L & 1 & 550 -- 1050 \\
NICMOS &G206 & 1 & 1400 -- 2500 \\ 
NICMOS &F166N & 2 & 1649 -- 1667  \\
NICMOS &F187N & 2 & 1864 -- 1884  \\
WFC3  & G141&2 & 1083 -- 1693$^*$ \\
IRAC &1 & 3 & 3200 -- 4000  \\
IRAC &2 & 2 & 4000 -- 5000  \\
IRAC &3 & 1 & 5100 -- 6500 \\
IRAC &4 & 7 & 6400 -- 9300 \\
MIPS & & 1 & 19500 -- 28500 \\ \hline
\end{tabular}\caption{List of the transit observations of HD 189733 from space used in the present study. (*:  only the edges of the wavelength range from the WFC3 data could be used, see \citet{gib12} for details.) }
\label{listobs}
\end{table}

\subsubsection{Limb-darkening}

Stellar limb darkening modifies the shape of a planetary transit, and therefore affects the radius determination. This is important when measuring the transmission spectrum, because limb darkening has a strong wavelength dependence. The effect of limb darkening on the measurement of the transmission spectrum gets stronger towards shorter wavelengths. It is negligible compared to other sources of uncertainty in the infrared, but becomes significant in the visible and UV. Fortunately, the HST data on HD 189733b is accurate enough to provide a constraint on limb darkening from the data itself, independently of stellar models. Some slight discrepancies between theoretical and observed limb darkening have been observed in STIS data for HD 209458b \citep{knu07}. \citet{hay12} have computed improved limb darkening coefficients from 3-D stellar atmosphere models, and find that these can capture the behaviour of the STIS data for HD 209458b better than the 1-D stellar atmosphere models used in \citet{knu07}.   For the temperature of HD 189733b, the difference between the prediction of 1-D and 3D models is lower, and both can account for the observed shapes of the transits with ACS and STIS. Here we use the new set of limb darkening coefficients from \citet{hay12}.

The remaining uncertainties on limb darkening coefficients is not a significant source of error for our combined transmission spectra.

\subsubsection{Instrumental effects}

In most of our data sets, the amplitude of residual instrumental systematics is similar to the signal expected from the planetary atmosphere. 
A careful consideration of the correction of instrument effects is therefore essential. The discussion of these corrections occupy a large part of the literature cited above, and we will only make a brief overview here.

Shortwards of 1 \micron\ (i.e. STIS and ACS) the instrumental systematics are relatively well-behaved.  They mainly result from the pointing drift and thermal expansion of the telescope, causing slight movements of the spectrum on the detector and changes of focus.
The telescope pointing and focus can be measured precisely on the data itself by monitoring the position and width of the spectrum on the detector. The correlation is easy to trace because of the very high signal-to-noise ratio of the flux measurements. As a result, in the case HD 189733, the uncertainties on these instrumental corrections are smaller than those due to the activity of the host star (see Section~\ref{starspots}), and than the signal of the atmosphere of the planet.

Further in the infrared, instrumental systematics become one of the dominant sources of uncertainty. The NICMOS and WFC3 data show fluctuations in the flux measurements that are thought to be correlated to changes in the temperature of the detector, and this temperature cannot be tracked as precisely as the pointing and focus changes. The \Spitzer\ measurements are affected by a Òramp effectÓ and a Òpixel phase effectÓ, that have been abundantly discussed in the relevant literature and limit the maximum accuracy of transit and eclipse measurements (none of these instruments had been originally designed for high-accuracy time series photometry of bright sources).

Among the infrared datasets for HD 189733b, the NICMOS {\it photometric} measurements are the ones least beset by these problems. Since they are passband-integrated photometric measurements rather than spectra, the instrumental effects have no wavelength-dependent component. The drawback is that the observations in different wavelengths are not simultaneous, rendering transmission spectrum measurements degenerate with changes in the star or in the planetary atmosphere. The NICMOS {\it grism} measurements are affected by time- and wavelength-dependent systematics, that are difficult to correct entirely because their relation with measured instrumental parameters is much looser than for ASC and STIS. Gaussian-process models of these effects show that linear decorrelation is not sufficient, and that non-linear relations between the measured instrumental parameters and their effect on the measured fluxes must be considered \citep[see][]{gib11,gib12b}.

In Spitzer measurements, the instrumental effects are generally larger than the atmospheric signatures. This has been a severe limitation in the context of the measurement of secondary eclipses and day-side surface brightness for many exoplanets. A much better understanding of these effects has now been reached \citep[see][for HD 189733]{des09,des11,knu12}. There are two effects, a gradual increase of the detector's sensitivity during the exposure, strongest in the shorter-wavelength channels, and a dependence on pointing, causing flux variations when the satellite pointing varies, because of intra-pixel sensitivity variations in the detector.

\subsubsection{Flux from the planet}

When converting transit depths into planet-to-star radius ratio, we consider the contribution of the emission from the night side of the planet to the total flux. We use a $T=900$~K blackbody distribution for the planetary emission (appropriate for the night side of the planet seen during transit), and a $T=5000$ K blackbody for the stellar emission. The relation between the observed and true radius ratio $r $ is:
\begin{eqnarray*}
r^2_{\rm obs} &=& \frac{R_{pl}^2}{R_*^2+R_{pl}^2 B^{T_{pl}}_\lambda/ B^{T_*}_\lambda} \\
&\simeq & \left(\frac{R_{pl}}{R_*}\right)^2 / \left(1+\left(\frac{R_{pl}}{R_*}\right)^2 \frac{B^{T_{pl}}_\lambda}{B^{T_*}_\lambda}\right) \\
&=& r_{\rm true}^2 (1- r_{\rm true}^2 \frac{B^{T_{pl}}_\lambda}{B^{T_*}_\lambda})
\end{eqnarray*}

This correction amounts to $1.2\cdot 10^{-4}$ on the radius ratio  at 8 \micron, $0.27 \cdot 10^{-4}$ at 3.6 \micron. It is negligible in the visible, and smaller than the other sources of uncertainty at all wavelengths. As neither the planet nor the star are perfect blackbody radiators, these values are indicative (stellar models can be different in flux from a blackbody by $\approx$ 20\%).

\section{Impact of stellar variability and star spots}

\label{starspots}

\begin{figure}
\resizebox{8cm}{!}{\includegraphics{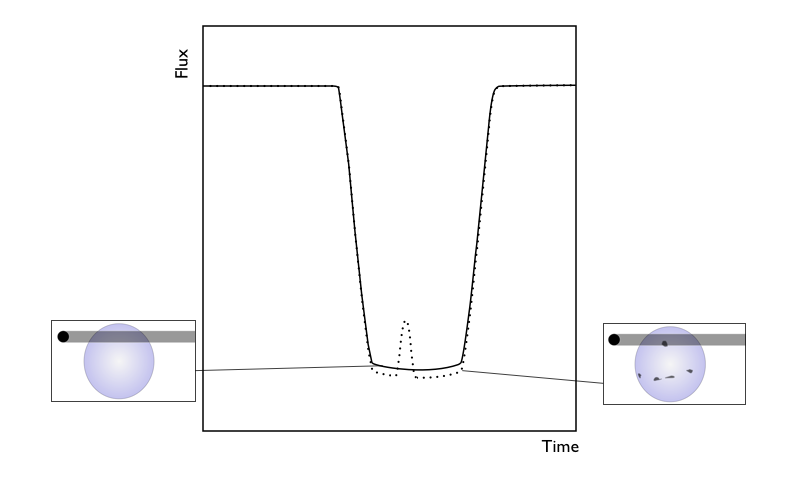} }
\caption{ Effect of stellar spots on a planetary transit light curve. Spots not occulted by the planet produce a deeper transit. Spots occulted by the planet produce flux rises on the timescale of the transit ingress/egress.}
\label{spot1}
\end{figure}

HD 189733 is an active K dwarf. Star spots modulate the total brightness of the star along the $\sim$12-days rotation cycle of the star \citep{win07}. Coincidentally, the dimming from star spots on HD 189733 in the visible, 1-2 \%, is comparable to the dimming produced by the planetary transit. Accounting for the presence of star spots when calculating the transmission spectrum is therefore essential. Figure \ref{spot1} illustrates the effect of star spots on transit depth measurements. For a planetary transit across an unspotted star, the depth of the flux dimming is proportional to the fraction of the stellar disc occulted by the planet, i.e. to the square of the radius ratio:

\[
d \propto \frac{A_{pl}}{A_*} =  \frac{R_{pl}^2}{R_{*}^2}
\] 

If the surface of the star is spotted, then the relation becomes:
\begin{equation}
d \propto \frac{\alpha R_{pl}^2}{ \alpha' R_{*}^2}
\label{alphas}
\end{equation}

where $\alpha$ is the mean brightness of the part of the star occulted by the planet,  and $\alpha '$ is the mean brightness of the stellar surface compared to a spot-free equivalent.

Thus there are two distinct ways for star spots to affect the recovered radius ratio: 
\renewcommand{\labelitemi}{$-$}
\begin{itemize}
\item first, star spots occulted by the planet during the transit reduce the transit depth, leading to an {\it underestimation} of the transit radius by a factor $\sqrt{\alpha}$.

\item second, star spots not occulted by the planet but situated on the side of the star visible during the transit will  lead to an {\it overestimation} of the planetary radius by a factor $\sqrt{\alpha'}$.

\end{itemize}

Both effect have a similar amplitude - they are proportional to the fraction of the stellar flux blocked by star spots.Both effects are also wavelength-dependent, because star spots have a different temperature and a different spectral energy distribution than the rest of the stellar disc, so that they modify not only the recovered radius ratio, but also the inferred transmission spectrum. For more quantitative details see \citet{pon08} and \citet{sin11}.


The present description neglects two further effects of variability on the transit light curve: the limb darkening of the spots, and the effect of brighter active regions on the star, faculae and plages. The first effect is negligible to our level of accuracy. The second could be important, but none of the space and ground-based monitoring of the HD~189733b system available shows a significant signature of the crossing of a brighter region. This is coherent with our understanding of the variability of cool stars and the Sun. Dark spots are well-defined, large regions of lower flux, while faculae and plages are distributed more evenly across the stellar surface and tend to average out in the global flux. 
 
To recover the planetary radius from the transit data, we must estimate $\alpha$ and $\alpha'$ for each transit, including their wavelength dependence.
 
Fortunately, a large body of data has now been gathered on the variability of HD 189733 and the characteristics of its spots, and we are in a position to introduce corrections for these effects and assess their uncertainty in a much more solid way than was possible in earlier studies. 

We describe below how the spot corrections are calculated and implemented\footnote{We do not apply spot corrections to the MIPS data at 24 \micron, since they are negligible for that dataset compared to other sources of uncertainty.}. It is necessary to delve in some details, since they constitute a dominant source of uncertainty on the final transmission spectrum.




\subsection{Correction for spots not crossed by the planet}

Spots on the visible surface of the star affect the transmission spectrum by causing the occulted and unocculted parts of the star to have different spectra. This is the factor $\alpha'$ in Equation~\ref{alphas}.  

We assume that the mean spectrum of the spots does not change with time. Then:

\begin{equation}
\alpha' = 1 - f(t)  c_\lambda
\end{equation}
where $f(t)  $ is the flux dimming due to spots at some reference wavelength $\lambda_0$, and $c_\lambda$ the ratio of the effect of spot between $\lambda$ and the reference wavelength $\lambda_0$.

\subsubsection{Estimating $f(t)$}

The factor $f(t)$ is estimated, as in previous studies, by monitoring the variability of HD 189733. We benefit from an almost continuous photometric monitoring of the star with the APT photometer \citep{hen99} over more than 5 years to measure the level of unocculted spots during the HST observations. We use Gaussian processes to interpolate the variability data in time, as detailed in Appendix A. Gaussian processes allow a Bayesian interpolation of the data with a minimum of assumptions on the functional form and regularity of the light curve, and are especially tailored to calculate realistic uncertainties. The photometric data and interpolation is shown in Fig.~\ref{GP}.
The resulting factors $f(t)$ at the epochs corresponding to space observations are given in Table \ref{dimming}. The reference wavelength is that used in the APT photometry, the mean of the $b$ and $y$ Str\"omgren filters (approx. 4500-4900 \AA\ and 5300-5700 \AA\ respectively).

\begin{table}
\begin{tabular}{lrrr}
\hline
Instrument & Date& $\Delta f$ & $\sigma_{f}$ \\
 & [BJD-2450000] & & \\
 \hline
ACS (G800L) & 3877.20 & 0.0021 & 0.0012  \\
WFC3 (G141) & 5510.09  & $-$0.0037 & 0.0010  \\
WFC3 (G141) & 5443.52 & $-$0.0022 & 0.0046  \\
NICMOS (F166N) & 4589.36  & 0.0033 & 0.0033  \\
NICMOS (F187N) & 4571.65 &  0.0054 & 0.0016  \\
NICMOS (G206) & 4219.98 & 0.0100 & 0.0033  \\
Spitzer (IRAC1) & 4429.68  & $-$0.0016 & 0.0029  \\
Spitzer (IRAC1) & 4039.22 & 0.0147 & 0.0018  \\
Spitzer (IRAC1) & 5559.55 & $-$0.0093 & 0.0045  \\
Spitzer (IRAC2) & 4427.47 & $-$0.0003 & 0.0028  \\
Spitzer (IRAC3) & 4429.68  & $-$0.0016 & 0.0029  \\
Spitzer (IRAC4) & 4281.00  & 0.0027 & 0.0035  \\ \hline
\end{tabular}
\caption{Values of stellar flux variation used in the spot corrections, calculated from the Gaussian-process interpolation of the APT photometry. Note: the stellar flux could not be derived for the STIS measurement. These were connected to the ACS measurements using the overlap in wavelength (see text).}
\label{dimming}
\end{table}

\begin{figure*}
\resizebox{18cm}{!}{\vspace{-2cm}\includegraphics{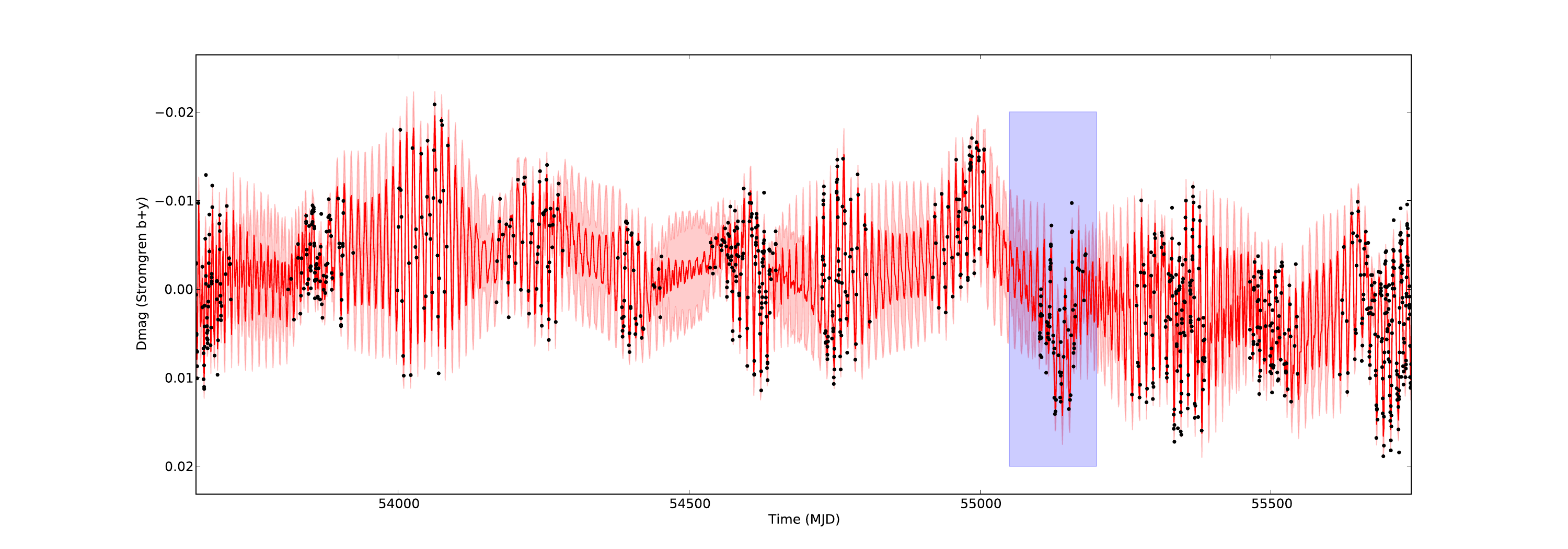}}
\caption{Flux measurements for HD 189733 collected with the APT photometer over six years, and Gaussian process interpolation. The red line shows the mean value, and the pink area the 1-sigma contour of the possible interpolations. The rectangular shading identifies the poorer observing season. The interpolation process is described in Appendix A. }
\label{GP}
\end{figure*}


The ground-based light curve and its Gaussian-process interpolation produce an estimate of $f(t)$ relative to an arbitrary reference level. To translate that into a  value of $f(t)$ that can be used to correct the spectroscopic data, it is also necessary to know the level of flux corresponding to an entirely spot-free surface. This reference level cannot be measured directly. It does not correspond to the maximum of the light curve, since it is possible -- indeed likely -- that even at its maximum flux the visible surface of the star is still affected by spots.

Fortunately there are several ways to estimate this reference level indirectly, and several lines of evidence suggest a value of the order of 1-2 \% of the flux. 

\begin{enumerate}
\item The statistics of spot-crossing events during the HST observations suggests a 1-2 \% reference level  \citep[see][for details]{sin11}. This is valid if the portion of the star crossed by the planet is typical of the whole stellar surface in terms of spot coverage.

\item \citet{aig12} find that in stochastic star spot simulations as well as in the Sun, the reference level is above the maximum of the light curve by a value that is comparable to the variance of the light curve itself. A much higher value requires unnatural spot configurations, and in the case of HD 189733 these configurations would have to survive for many spot cycles in order to maintain the observed variance of the light curve much below the reference level.

\item The transmission spectrum in the visible does not show any spectral feature associated with star spots, most notably the broad wings of the stellar sodium line and the MgH line, which have a steep temperature dependence \citep[see][ figure 10]{sin11}. This would be expected if the reference level was much higher than 1-2 percent.
\end{enumerate}

\subsubsection{Estimating $c_\lambda$}

To estimate $c_\lambda$, we use the same procedure as \citet{pon08} and \citet{sin11}. As far as can be gathered from the spectrum changes during the spot crossings observed with STIS and ACS, a mean temperature of $4250 \pm 250$ K provides a good model of the spectral energy distribution of spots.

We model the spot-free photosphere using a MARCS stellar model \citep{gus08} with solar metallicity, log g = 4.5 and Teff = 5000 K. For the spots, we used similar models, but with cooler temperatures (4000, 4250 and 4500 K). We refer to each spectrum by its temperature, for example F4000($\lambda$) is the 4000 K spectrum. \def\f{\delta}
We compute a model spectrum for a star with a fraction $\f$ of the projected visible area covered in spots at temperature T:

$F_{spotted}(\f, \lambda, T ) = (1-\f)F_{5000}(\lambda) + \f FT (\lambda)$	
where we have ignored limb-darkening.

We then compute the resulting flux relative to the spot-free case for a given observational setup:

\begin{equation}
 f_i(\f,T) = \frac{ \int F_{spotted}(\f,\lambda,T)L_i(\lambda)d\lambda }{\int F_{5000}(\f,\lambda,T)L_i(\lambda)d\lambda}
 \end{equation}
 
where $L_i(\lambda)$ is the combined instrument, filter and detector throughput for observational setup $i$. Where available, we used resolved transmission curves, linearly interpolated to the resolution of the MARCS spectra. Otherwise we used top-hat  functions for the relative spectral response curve (in W/nm), specifying only the minimum and maximum wavelengths.
 
To combine the out-of-transit photometry from different observatories, we need the ratio between the amplitude of the brightness variations in two observational setups. This is given by:

\begin{equation}
R_{ij}(\f,T)=  \frac{1-f_i(\f,T)}{1-f_j(\f,T)}
\end{equation}

We report these ratios in Table \ref{factor}, where $i$ is always the APT. Note that the value of $\f$ has minimal impact on the amplitude ratios, provided $\f<< 1$. We used $\f$ = 0.01 to obtain the ratios reported in the table.


\begin{table}
\begin{tabular}{lcrrr}
\hline
Instrument & Wavelength(nm) & $c_{4000}$ & $c_{4250}$ & $c_{4500}$\\
\hline
STIS (G430L) & 290 -- 370 &  1.130 &  1.244 &  1.403 \\
STIS (G430L) & 370 -- 420 &  1.082 &  1.160 &  1.255 \\
STIS (G430L) & 420 -- 470 &  1.049 &  1.072 &  1.106 \\
STIS (G430L) & 470 -- 520 &  1.037 &  1.065 &  1.087 \\
STIS (G430L) & 520 -- 570 &  0.965 &  0.955 &  0.946 \\
STIS (G750M) & 581 -- 592 &  0.958 &  0.930 &  0.902 \\
STIS (G750M) & 592 -- 603 &  0.936 &  0.897 &  0.863 \\
STIS (G750M) & 603 -- 615 &  0.908 &  0.880 &  0.856 \\
STIS (G750M) & 615 -- 626 &  0.947 &  0.908 &  0.864 \\
STIS (G750M) & 626 -- 637 &  0.917 &  0.881 &  0.845 \\
STIS (G750M) & 588 -- 590 &  1.047 &  1.067 &  1.049 \\
STIS (G750M) & 589 -- 589 &  1.086 &  1.168 &  1.230 \\
ACS (G800L) & 550 -- 600 &  0.948 &  0.922 &  0.898 \\
ACS (G800L) & 600 -- 650 &  0.916 &  0.884 &  0.850 \\
ACS (G800L) & 650 -- 700 &  0.882 &  0.849 &  0.808 \\
ACS (G800L) & 700 -- 750 &  0.811 &  0.789 &  0.760 \\
ACS (G800L) & 750 -- 800 &  0.762 &  0.746 &  0.721 \\
ACS (G800L) & 800 -- 850 &  0.724 &  0.714 &  0.691 \\
ACS (G800L) & 850 -- 900 &  0.702 &  0.692 &  0.664 \\
ACS (G800L) & 900 -- 950 &  0.675 &  0.665 &  0.638 \\
ACS (G800L) & $\,\,$950 -- 1000 &  0.660 &  0.647 &  0.619 \\
ACS (G800L) & 1000 -- 1050 &  0.653 &  0.636 &  0.603 \\
WFC3 (G141) & 1099 -- 1153 &  0.601 &  0.575 &  0.545 \\
WFC3 (G141) & 1500 -- 1694 &  0.484 &  0.338 &  0.300 \\
NICMOS (G206) & 1463 -- 1520 &  0.525 &  0.408 &  0.365 \\
NICMOS (G206) & 1520 -- 1578 &  0.504 &  0.365 &  0.324 \\
NICMOS (G206) & 1578 -- 1635 &  0.470 &  0.320 &  0.283 \\
NICMOS (G206) & 1635 -- 1693 &  0.459 &  0.299 &  0.263 \\
NICMOS (G206) & 1693 -- 1750 &  0.449 &  0.290 &  0.255 \\
NICMOS (G206) & 1750 -- 1807 &  0.447 &  0.289 &  0.254 \\
NICMOS (G206) & 1807 -- 1865 &  0.441 &  0.286 &  0.250 \\
NICMOS (G206) & 1865 -- 1922 &  0.431 &  0.280 &  0.244 \\
NICMOS (G206) & 1922 -- 1980 &  0.447 &  0.297 &  0.254 \\
NICMOS (G206) & 1980 -- 2037 &  0.436 &  0.289 &  0.250 \\
NICMOS (G206) & 2037 -- 2094 &  0.418 &  0.276 &  0.241 \\
NICMOS (G206) & 2094 -- 2152 &  0.411 &  0.274 &  0.241 \\
NICMOS (G206) & 2152 -- 2209 &  0.397 &  0.267 &  0.235 \\
NICMOS (G206) & 2209 -- 2267 &  0.393 &  0.270 &  0.238 \\
NICMOS (G206) & 2267 -- 2324 &  0.388 &  0.278 &  0.251 \\
NICMOS (G206) & 2324 -- 2381 &  0.391 &  0.293 &  0.267 \\
NICMOS (G206) & 2381 -- 2439 &  0.392 &  0.303 &  0.278 \\
NICMOS (G206) & 2439 -- 2496 &  0.404 &  0.305 &  0.272 \\
NICMOS (F166N) & 1649 -- 1667 &  0.455 &  0.298 &  0.263 \\
NICMOS (F187N) & 1864 -- 1884 &  0.423 &  0.271 &  0.236 \\
Spitzer (3.6) & 3200 -- 3900 &  0.369 &  0.272 &  0.231 \\
Spitzer (4.5) & 4000 -- 5000 &  0.283 &  0.253 &  0.230 \\
Spitzer (5.8) & 5100 -- 6500 &  0.266 &  0.246 &  0.224 \\
Spitzer (8.0) & 6400 -- 9300 &  0.267 &  0.242 &  0.210 \\\hline
\end{tabular}\caption{The scaling factors for the effect of star spots at three different temperature, 4000, 4250 and 4500 K. }
\label{factor}
\end{table}

\subsubsection{Discrepant APT season}

\begin{figure}
\resizebox{8cm}{!}{\includegraphics{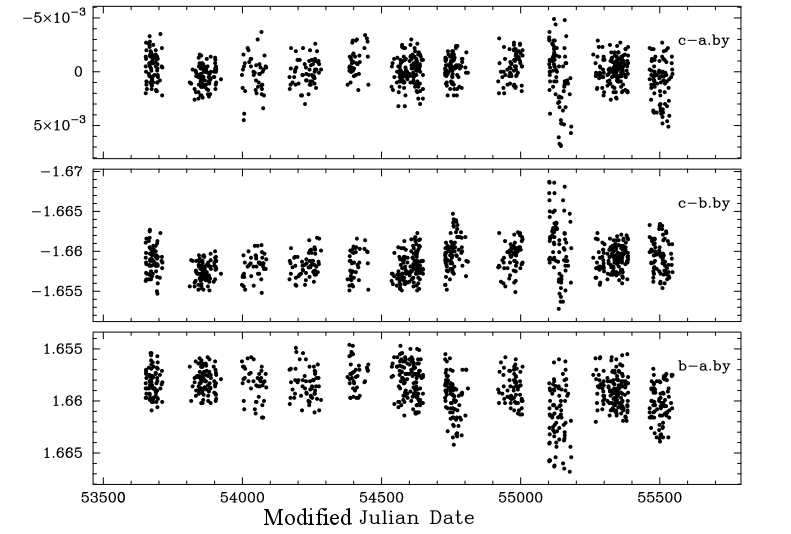}}
\caption{Time series of the difference in flux between the three comparison stars in the APT photometry. Note the noisier season around MJD=55150.}
\label{compphot}
\end{figure}

Fig. \ref{compphot} shows the APT photometry for  the three comparison stars relative to each other. This suggests that one of the seasons is of markedly lower quality than the others (between JD 2455150 and 2455400) and may have an incorrect zero-point for HD 189733. This corresponds to the season with the most discrepant behaviour of the Gaussian process interpolation for HD 189733 (Fig. \ref{GP}). The Gaussian process approach does not include a model of a zero-point change (which would be completely degenerate with a change in flux of HD 189733). We therefore choose not to use the data for this season. 

Only the STIS spectrum was taken during this season, and indeed, if we re-scale the STIS and ACS observation using the Gaussian process interpolation, we observe a large discrepancy between the two spectra over the wavelength range at which they overlap. Since we do not consider actual change in atmospheric properties at this level (``weather'') to be plausible, we suspect higher uncertainties in the measurement of the stellar flux during the lower-quality season to be responsible for the mismatch.

Like in \citet{sin11}, we therefore use the assumption that the spectrum has to be compatible across the overlap in wavelength and choose the spot level that correctly connects the STIS and ACS spectra.

 \subsection{Corrections for spots crossed by the planet}

Occultations of star spots by the planet are clearly seen in the ACS and STIS transit curves.  Figures \ref{resvis} and \ref{resir} show the residuals around the best-fit transit model (with the same system parameters for all data sets), for visible and infrared data respectively\footnote{Throughout this paper we designate as "visible" the observations shortwards of 1\micron, and "infrared" beyond. This does not correspond to human eye sensitivity, but rather to the use of CCD-detectors for the first category and infrared-array detector for the second, which separates the observations into different categories for the purpose of reduction and analysis.}. One remarkable feature of the ensemble data is that every single visit in the visible shows the signature of occulted star spots. No such events are seen in the infrared.  The signal-to-noise ratio of spot occultations is much lower in the infrared data, both because the noise on individual data points is higher, and because the spot-to-photosphere contrast becomes lower at longer wavelengths. This strongly suggests that undetected spot crossings are present in the infrared data as well. Figure \ref{residus} shows quantitatively how star spots can be missed in the infrared data.

\begin{figure*}
\resizebox{15cm}{!}{\includegraphics[angle=90]{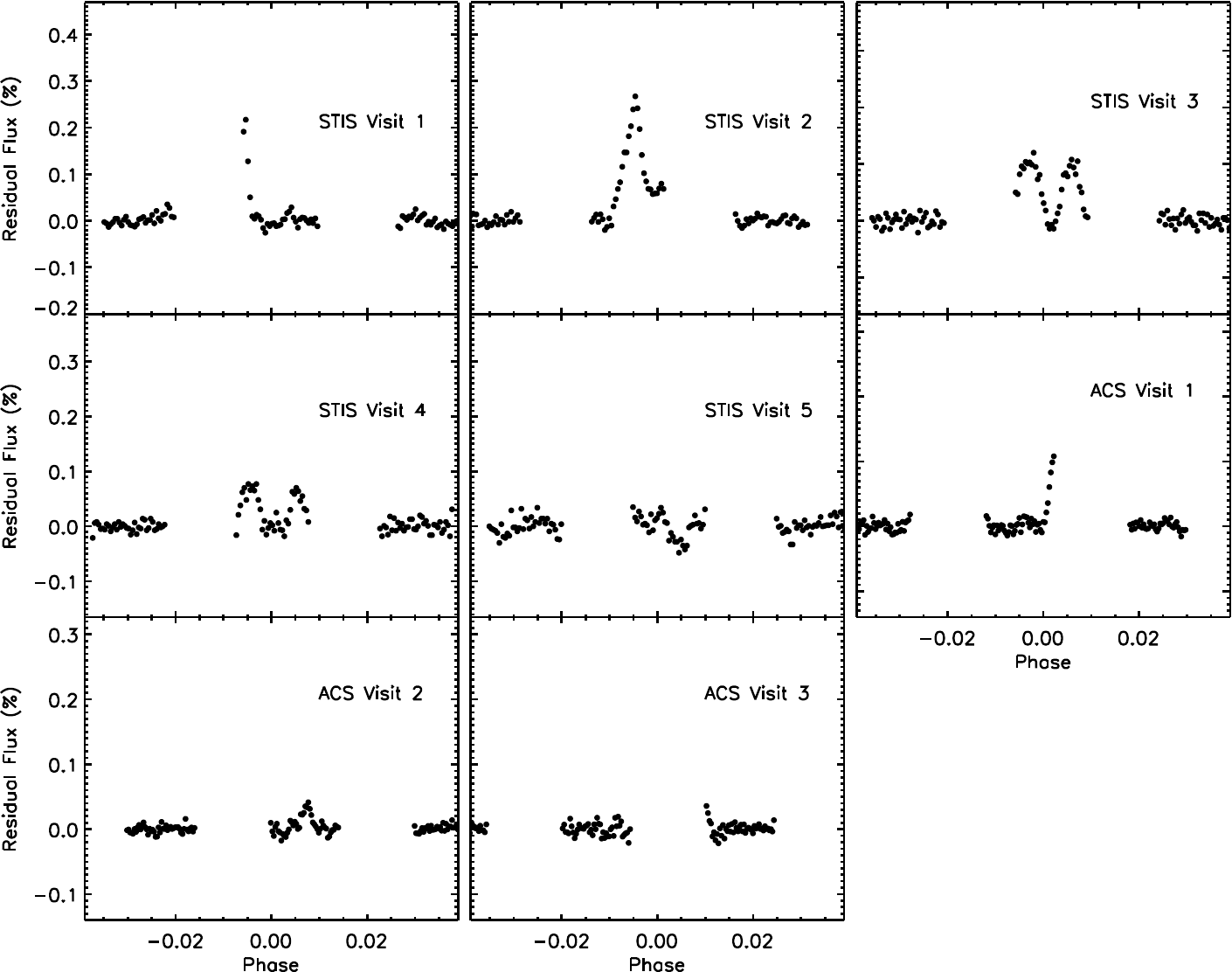} }
\caption{ Residuals compared to the transit model for the STIS and ACS visits, integrated over all wavelengths. The vertical axes are scaled so that the same star spot would produce an effect of the same amplitude (using the wavelength scaling in Table \ref{factor} with $T_{\rm spot}$ = 4250~K).}
\label{resvis}
\end{figure*}

\begin{figure*}
\resizebox{9.6cm}{!}{\includegraphics[angle=90]{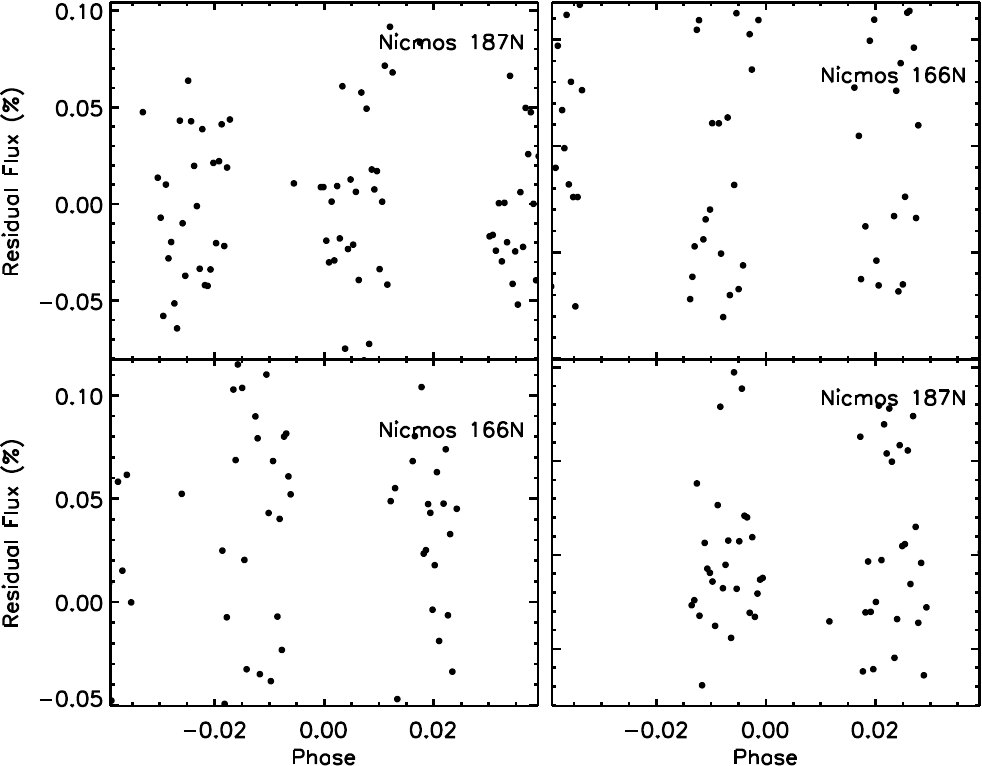} }
\caption{ Residuals compared to the transit model for the NICMOS photometric visits, integrated over all wavelengths. The data was binned by groups of 7 points to make the temporal coverage comparable to the ACS and STIS data of Fig. \ref{resvis}. The vertical axis are scaled so that the same star spot would produce an effect of the same amplitude (using the wavelength scaling in Table \ref{factor} with $T_{\rm spot}$ = 4250~K).}
\label{resir}
\end{figure*}


At visible wavelengths, we use only the parts of the data that appear unaffected by significant star spot crossings. This corresponds to using $\alpha = 1$ in Equation \ref{alphas}.

\begin{figure*}
\resizebox{16cm}{!}{\includegraphics{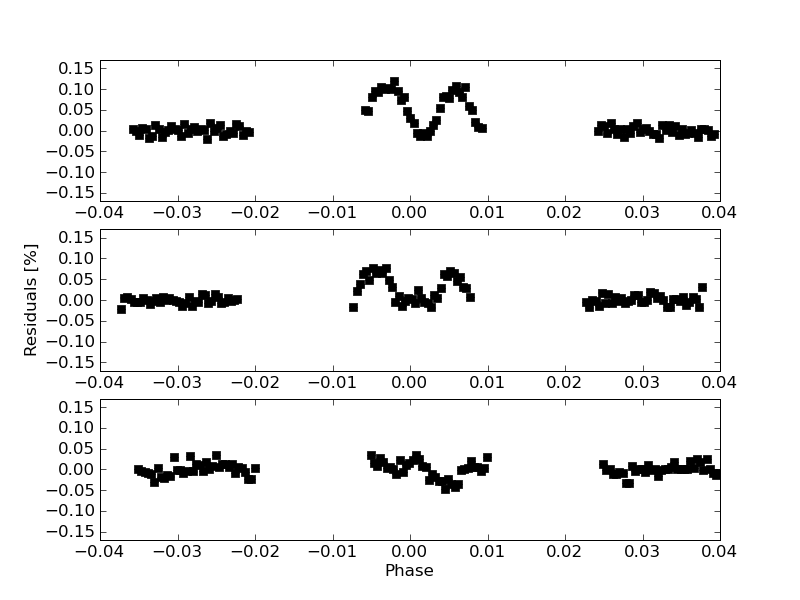} \includegraphics{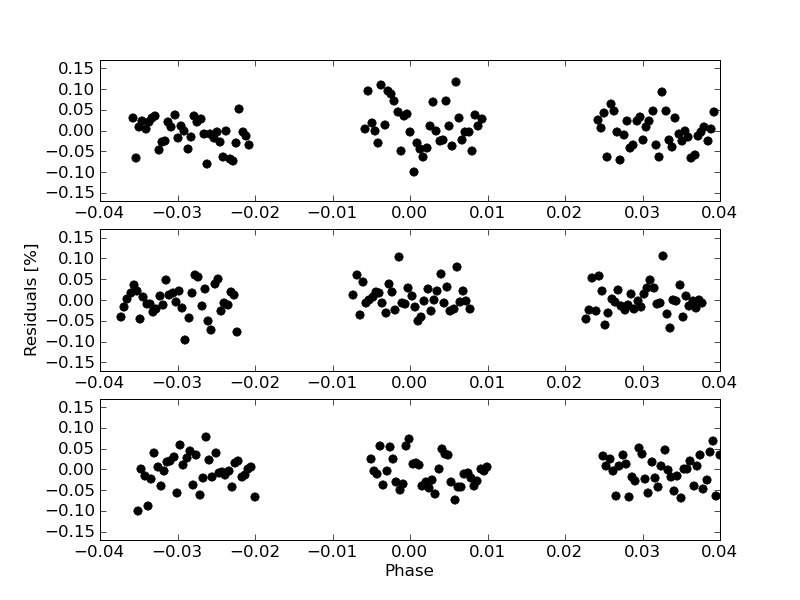}}
\caption{ {\bf Left:} residuals around a transit model from the three medium-resolution STIS HST visits, showing the ubiquitous effect of occulted spots. {\bf Right:} Same residuals, with the spot effect scaled to the expected amplitude at 8 \micron, and added errors corresponding to those of \sst\ measurements at 8 \micron\ ($3.7 \times 10^{-4}$ per minute).}
\label{residus}
\end{figure*}

 In the infrared, since individual spot crossings cannot be identified, we must rely on an estimate of the average effect of occulted spots. If the spots are randomly distributed on the surface of the star, the effect of occulted and unocculted spots will compensate each other on average ($\alpha = \alpha '$ on average). The effect of spot crossing by the planet will be to increase the error in this estimate, due to the low-number statistics of individual spot crossings.  
 
 We estimate the dispersion introduced by unrecognised spot crossings in the infrared in three ways: first using the statistics of spot crossings in visible data, second using the statistics of repeated depth measurements in the infrared in the same passbands, third, using the variability in the light curve of the star:

 \begin{enumerate}
 
\item  the standard deviation of the depth measured in the ACS and STIS data including the spots (see Figs. \ref{resvis}) is 314$\cdot 10^{-6}$. This, while obviously affected by low-number statistics, is based on enough transits to be a meaningful estimate of the average effect of spot crossings. Scaled with the contrasts of Table \ref{factor}, using T=4250 K for the spots, it amounts to a scatter of 94$\cdot 10^{-6}$ on the transit depth near 1.6 \micron\ (NICMOS F116N filter) and a scatter of 75$\cdot 10^{-6}$ for the Spitzer 8 \micron\ channel. The last number translates into 0.3 \% of the total transit depth.

\item  Agol et al. (2008) measured a residual variations of 0.6~\% in the measured transit depth for seven measurements at 8 \micron\ with Spitzer. These authors identify spot crossings as a likely dominant source of this scatter, as they note that the residuals do not correlate with the stellar flux, indicating that the unocculted spots cannot entirely explain this scatter (they also argue that random error are smaller than this scatter).

\item  The total flux of the star in the blue (Str\"omgren $b$ and $y$ filters) varies by 1-3 \%, depending on the seasons. If the planet does not cross a special latitude respective to the position of the spots, then star spots amounting to a 1-3\% dimming in the APT $b+y$ passband on the path of the planet will scale to a 0.24-0.72~\% effect on the depth of a transit at 8 \micron.

\end{enumerate}

These three independent estimates are compatible with each other, given that the second is the sum of the contribution of the occulted stat spots with the effect of unocculted star spots (of the same amplitude on average) and the instrumental systematics.  A scatter of 0.3~\% at 8 \micron\ on $\alpha$ due to unrecognised star spot crossings is compatible with all three.

We therefore use 0.3 \% at 8 \micron\ as an additional uncertainty in the depth measurement of individual transits due to unidentified spot crossings, and use $\alpha=\,<\!\!\alpha'\!\!>$ to calculate our best estimate of the transit radius in the infrared data. This value is scaled with the factors in Table \ref{factor} at other wavelengths.



\subsection{Limitations of the spot corrections}

Several lines of evidence and cross-checks allow us to build some confidence on the spot corrections. We have also stayed as conservative as possible in our assumptions about the effect of unseen spots in the infrared data. As a results, the uncertainties in the infrared measurements are significantly larger than in previous studies.

Nevertheless, some coincidences and compensating effects remain possible. We identify some of them here.

The mean effective temperature of spots is calculated from the large spot crossings in the ACS and STIS data. It is possible that this temperature is only representative of larger spots, and that there is a large population of smaller spots with a weaker temperature difference. Such a "leopard skin" model for HD 189733 would modify the transmission spectrum in ways that would be virtually impossible to correct with the available data. Nevertheless, with smaller and more numerous spots, the effects of occulted and unocculted spots tend to average out over the scale of a full transit.

\begin{figure}
\resizebox{8cm}{!}{\includegraphics{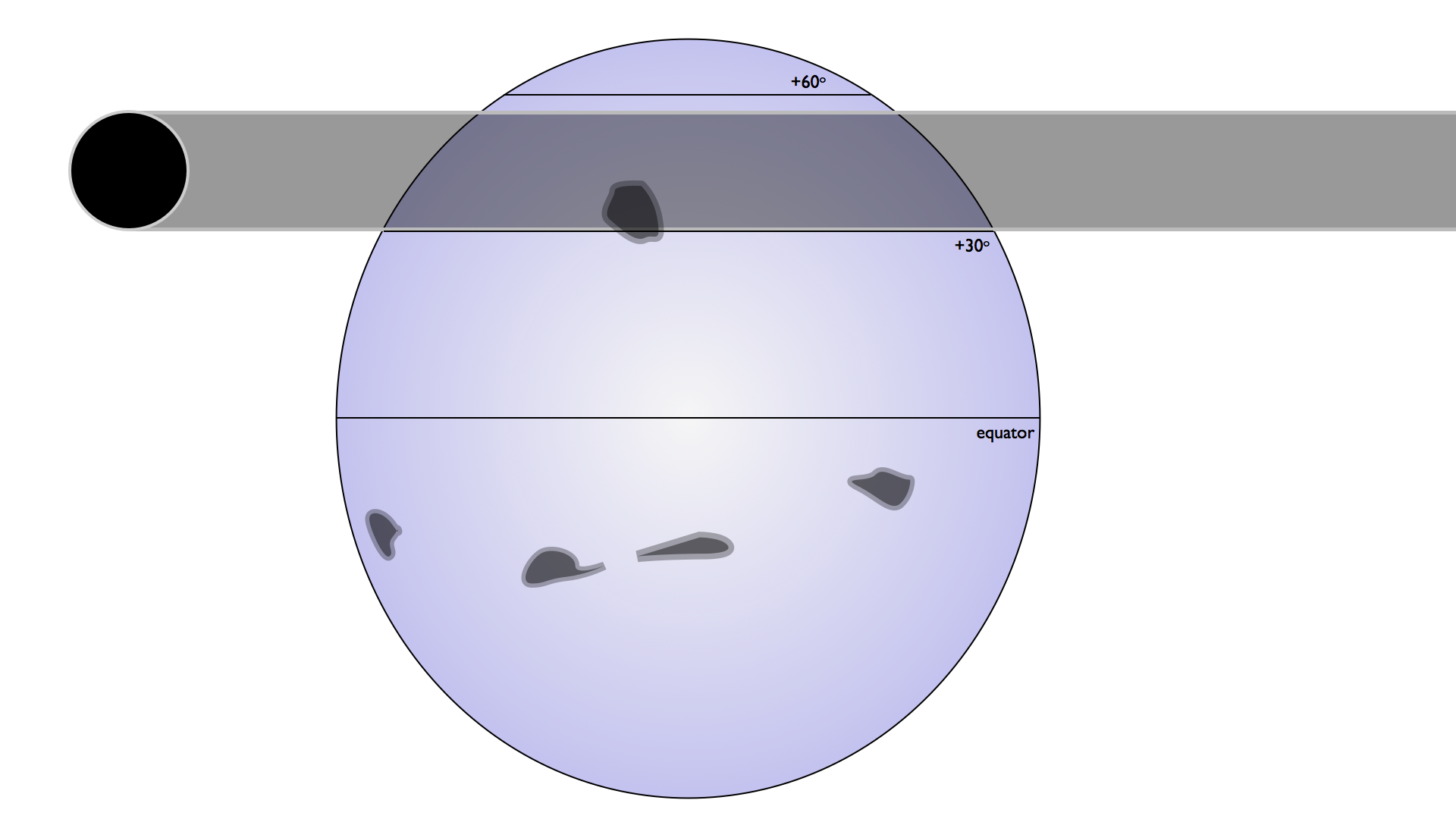} }
\caption{Geometry of the HD189733 system.}
\label{spot3}
\end{figure}

The level of unocculted spots, that we have estimated to be 1-2\% in the previous section, could be much higher. This scenario requires a certain number of coincidences to remain compatible with the observations. First, the spots must be preferentially situated out of the path taken by the planet across the star during the transit (the planet crosses the 31-55$^0$ latitude range of the star, see Fig.~\ref{spot3}), so as to reproduce the statistics of crossed spots and the variability of measured depth at 8 \micron. This is possible with a large polar spot region, for instance. The spot configuration would also have to remain remarkably stable, to reproduce the variability of the APT light curve over more than five years. Also, the MgH feature expected with such a high spot dimming \citep[see Fig.~10 of][]{sin11} is not seen, which requires the spot spectrum to be anomalous and different from a cooler stellar photosphere (this last point is not as unlikely as it sounds, the TrES-1 spot occultation observed with ACS in \citet{rab09} for instance show a spectrum for the spot that seems flatter than expected, and in our STIS spectrum, the expected MgH spectral feature of the spot near 5000 \AA\ seems  absent).

The distribution of spots could also be strongly uneven in latitude, undermining the $\alpha =\,<\!\!\alpha'\!\!>$ relation used for the infrared transit. This would bias the connection of the infrared and visible parts of the transmission spectrum.

We consider these situations less likely than our default assumptions, and will leave them as possible caveat unless they are supported by empirical evidence.

\subsection{``Weather'' variability}

\label{weather}

Reconstructing  a unique transmission spectrum from data taken at different times rests on the assumption that intrinsic variations in the planetary spectrum (``weather'') are smaller than the observational uncertainties.

Any variation between different runs at the same wavelength could be explained by actual variations in the transmission spectrum of the atmosphere of the planet. 
Nevertheless, in our most accurate datasets (ACS, STIS, NICMOS filters), radius measurements repeated with the same instrument agree with each other within the uncertainties without intrinsic variations in the planet (in the infrared, the uncertainties related to spot and instrument effect corrections are larger than the variations between different measurements). 

Moreover, the radius variations that we measure span several atmospheric scale heights, which is much larger than the expected variations due to Òplanetary weatherÓ.

We therefore assume that the observed variations are due to the mean atmospheric transmission spectrum around the planet combined with observational uncertainties, rather than the actual variations of the atmospheric transmission spectrum.

Building a single transmission spectrum from our data sets rests on this assumption.

%
%

\section{Results}

\begin{table*}
\begin{tabular}{lcccrr}
\hline
Instrument & BJD & Central & Band  & $R_p/R_s$  & $\sigma_{R_p/R_s}$  \\
(setting)visit  & [-2450000]  & Wavelength [\AA] & half-width [\AA] &  &  \\
 \hline
STIS(G430L)1    &    5155.12490    &    3300   &    400   &    0.15866    &    0.00043    \\   
STIS(G430L)2    &    5334.80977    &    3300   &    400   &    0.15734    &    0.00051    \\   
STIS(G430L)1    &    5155.12490    &    3950   &    250   &    0.15762    &    0.00020    \\   
STIS(G430L)2    &    5334.80977    &    3950   &    250   &    0.15732    &    0.00026    \\   
STIS(G430L)1    &    5155.12490    &    4450   &    250   &    0.15701    &    0.00024    \\   
STIS(G430L)2    &    5334.80977    &    4450   &    250   &    0.15728    &    0.00018    \\   
STIS(G430L)1    &    5155.12490    &    4950   &    250   &    0.15669    &    0.00023    \\   
STIS(G430L)2    &    5334.80977    &    4950   &    250   &    0.15706    &    0.00015    \\   
STIS(G430L)1    &    5155.12490    &    5450   &    250   &    0.15654    &    0.00023    \\   
STIS(G430L)2    &    5334.80977    &    5450   &    250   &    0.15672    &    0.00016    \\   
ACS(G800L)1    &    3877.20896    &    5750   &    250   &    0.15644    &    0.00014    \\   
STIS(G750M)    &    5148.46821    &    5865    &    57    &    0.15638    &    0.00027    \\   
STIS(G750M)    &    5148.46821    &    5895    &    11    &    0.15703    &    0.00011    \\   
STIS(G750M)    &    5148.46821    &    5980   &    57   &    0.15631    &    0.00022    \\   
STIS(G750M)    &    5148.46821    &    6095   &    55   &    0.15617    &    0.00036    \\   
STIS(G750M)    &    5148.46821    &    6207   &    57   &    0.15600    &    0.00027    \\   
STIS(G750M)    &    5148.46821    &    6321   &    57   &    0.15611    &    0.00019    \\   
ACS(G800L)1    &    3877.20896    &    6250   &    250   &    0.15610    &    0.00012    \\   
ACS(G800L)1    &    3877.20896    &    6750   &    250   &    0.15585    &    0.00011    \\   
ACS(G800L)1    &    3877.20896    &    7250   &    250   &    0.15572    &    0.00011    \\   
ACS(G800L)1    &    3877.20896    &    7750   &    250   &    0.15586    &    0.00012    \\   
ACS(G800L)1    &    3877.20896    &    8250   &    250   &    0.15552    &    0.00012    \\   
ACS(G800L)1    &    3877.20896    &    8750   &    250   &    0.15553    &    0.00013    \\   
ACS(G800L)1    &    3877.20896    &    9250   &    250   &    0.15546    &    0.00013    \\   
ACS(G800L)1    &    3877.20896    &    9750   &    250   &    0.15552    &    0.00016    \\   
ACS(G800L)1    &    3877.20896    &    10250   &    250   &    0.15496    &    0.00024    \\   
HST(WFC3)2    &    5510.09830    &    11050   &    230   &    0.15671    &    0.00084    \\   
HST(WFC3)1    &    5443.52370    &    11335   &    345   &    0.15549    &    0.00089    \\   
NICMOS(G206)    &    4219.98178    &    14920   &    285   &    0.15335    &    0.00129    \\   
NICMOS(G206)    &    4219.98178    &    15500   &    285   &    0.15330    &    0.00088    \\   
HST(WFC3)2    &    5510.09830    &    15925   &    785   &    0.15608    &    0.00055    \\   
HST(WFC3)1    &    5443.52370    &    16070   &    860   &    0.15543    &    0.00082    \\   
NICMOS(G206)    &    4219.98178    &    16070   &    285   &    0.15241    &    0.00078    \\   
NICMOS(F166N)1    &    4589.36894    &    16580   &    90   &    0.15516    &    0.00056    \\   
NICMOS(F166N)2    &    4611.60656    &    16580   &    90   &    0.15582    &    0.00080    \\   
NICMOS(G206)    &    4219.98178    &    16650   &    285   &    0.15405    &    0.00079    \\   
NICMOS(G206)    &    4219.98178    &    17220   &    285   &    0.15424    &    0.00098    \\   
NICMOS(G206)    &    4219.98178    &    17790   &    285   &    0.15511    &    0.00068    \\   
NICMOS(G206)    &    4219.98178    &    18370   &    285   &    0.15490    &    0.00070    \\   
NICMOS(F187N)1    &    4571.65062    &    18740   &    100   &    0.15459    &    0.00056    \\   
NICMOS(F187N)2    &    4689.26721    &    18740   &    100   &    0.15456    &    0.00048    \\   
NICMOS(G206)    &    4219.98178    &    18940   &    285   &    0.15572    &    0.00075    \\   
NICMOS(G206)    &    4219.98178    &    19510   &    285   &    0.15525    &    0.00069    \\   
NICMOS(G206)    &    4219.98178    &    20090   &    285   &    0.15370    &    0.00058    \\   
NICMOS(G206)    &    4219.98178    &    20660   &    285   &    0.15486    &    0.00080    \\   
NICMOS(G206)    &    4219.98178    &    21240   &    285   &    0.15395    &    0.00080    \\   
NICMOS(G206)    &    4219.98178    &    21810   &    285   &    0.15483    &    0.00068    \\   
NICMOS(G206)    &    4219.98178    &    22380   &    285   &    0.15490    &    0.00061    \\   
NICMOS(G206)    &    4219.98178    &    22960   &    285   &    0.15491    &    0.00073    \\   
NICMOS(G206)    &    4219.98178    &    23530   &    285   &    0.15432    &    0.00063    \\   
NICMOS(G206)    &    4219.98178    &    24110   &    285   &    0.15496    &    0.00070    \\   
NICMOS(G206)    &    4219.98178    &    24680   &    285   &    0.15520    &    0.00090    \\   
Spitz(IRAC1)    &    4429.68978    &    36000   &    4000   &    0.15471    &    0.00051    \\   
Spitz(IRAC1)    &    4039.22278    &    36000   &    4000   &    0.15547    &    0.00037    \\   
Spitz(IRAC1)    &    5559.55455    &    36000   &    4000   &    0.15452    &    0.00059    \\   
Spitz(IRAC2)    &    4427.47301    &    45000   &    5000   &    0.15538    &    0.00051    \\   
Spitz(IRAC2)    &    5189.05249    &    45000   &    5000   &    0.15543    &    0.00049    \\   
Spitz(IRAC3)    &    4429.68978    &    58000   &    7000   &    0.15476    &    0.00067    \\   
Spitz(IRAC4)    &    4281.00701    &    78500   &    14500   &    0.15510    &    0.00034    \\   
Spitzer(MIPS)    &    4398.60560    &    240000   &    45500   &    0.15459    &    0.00094    \\   \hline
\end{tabular}\caption{Transmission spectroscopy results for all data sets. Note that the uncertainties on the transit radius are not independent (i.e. the uncertainties on the differences between two values of the radius ratio measured with the same instrument at the same epoch is smaller than the combination of the uncertainties on the absolute radius ratios given above).}
\label{specresults}
\end{table*}

\label{secresults}

\subsection{Combined UV-to-IR transmission spectrum}

Table \ref{specresults} and Figure \ref{radii} give the resulting values for the planet size as a function of wavelength, after applying the spot correction and uncertainty estimates detailed in the previous Section, using $T_{\rm spots}= 4250 \pm 250$ K.

\begin{figure*}
\resizebox{14cm}{!}{\includegraphics{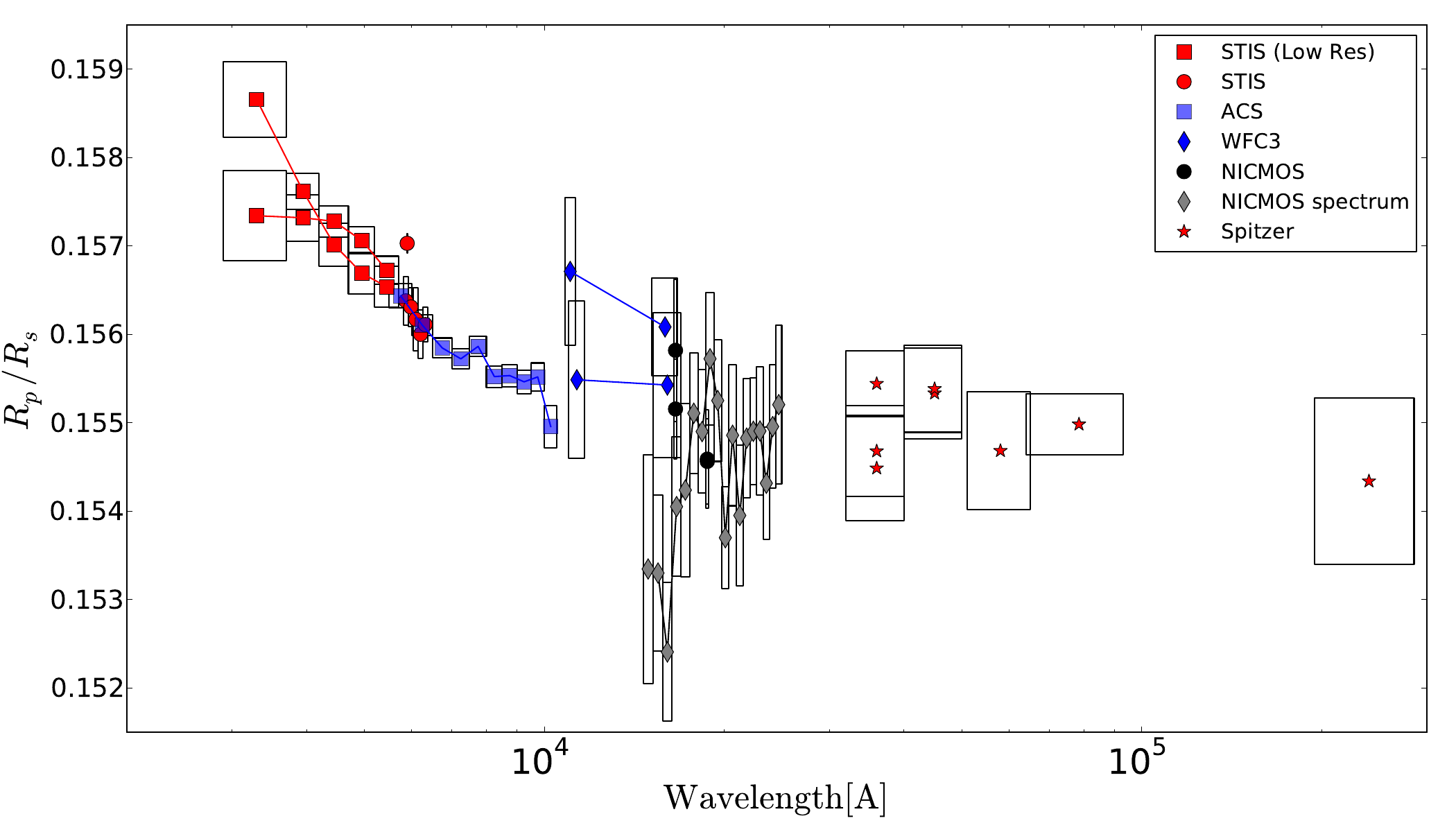}}
\caption{ Transmission spectrum data, with data sets and visits indicated separately. Lines connect values obtained at the same time with the same instrument.}
\label{radii}
\label{spectrum}
\end{figure*}

\begin{figure*}
\resizebox{12cm}{!}{\includegraphics{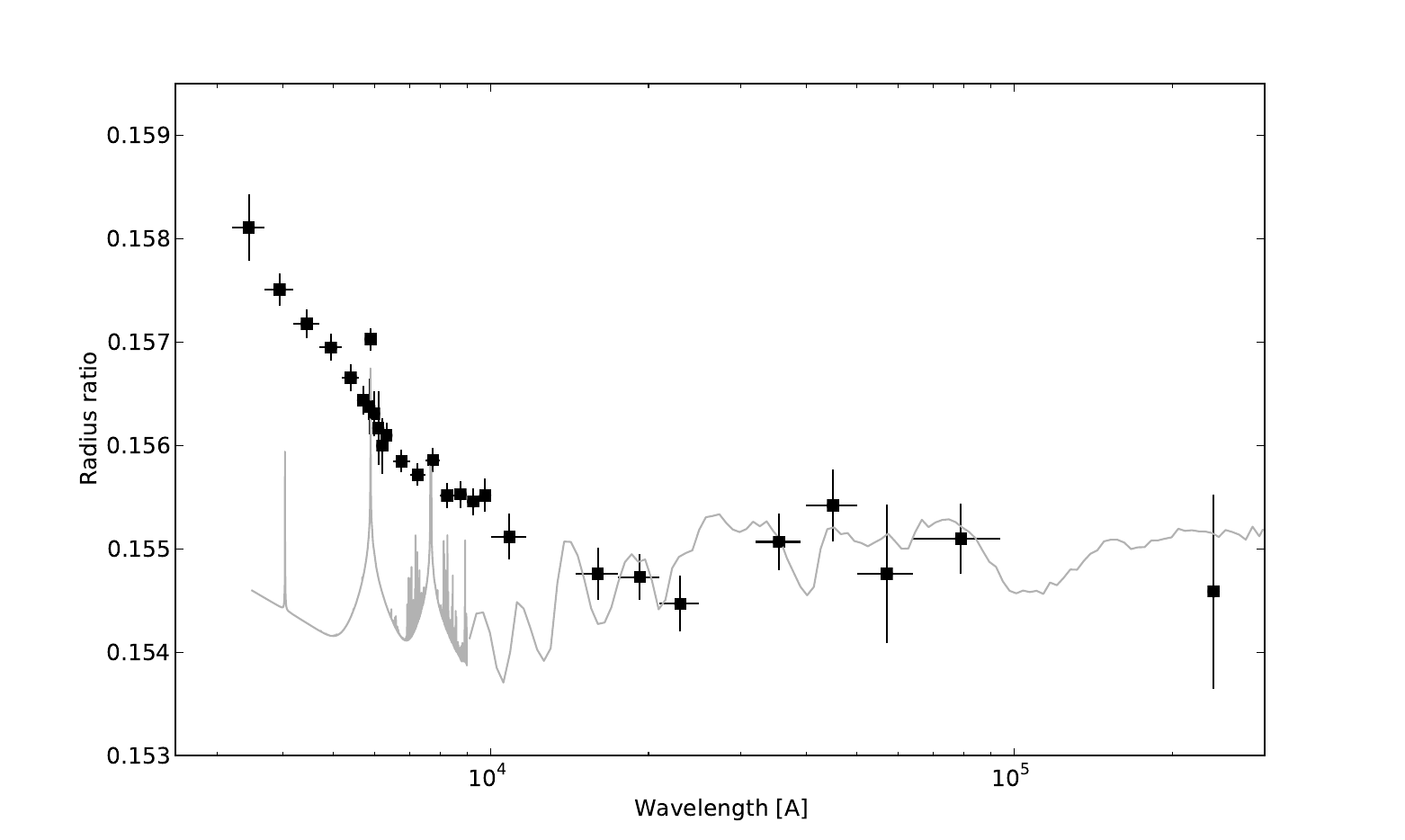}}
\caption{ Our combination of the available data into a single set of constraints, assuming that the intrinsic transmission spectrum of the planet is constant with time without our uncertainties. The grey line shows a synthetic spectrum with a dust-free model.  }
\label{binspectrum}
\end{figure*}

In Table \ref{binned} and Figure \ref{binspectrum}, the spectrum data is averaged over a limited number of passbands, arbitrarily chosen when combining data from different instruments. A condensate-free model of the transmission spectrum of HD 189733b from \citet{for10} is plotted for comparison.

The passband-average data in Table \ref{binned} are given for comparison with broad-brush models of the dependence of transit radius on wavelength. Detailed models of the transmission spectrum should be compared to the full data in Table \ref{specresults} rather than the binned version, adding, if necessary, the detailed shape of the sodium line reported in \citet{hui12}.

 \begin{table}
\begin{tabular}{cr} \hline
Passband &  Radius ratio \\
 (nm)   \\ 
 \hline
320-370  & 0.15811 $\pm$ 0.00032\\ 
370-420  & 0.15751 $\pm$ 0.00016\\
420-470  & 0.15718 $\pm$ 0.00014\\
470-520  & 0.15695 $\pm$ 0.00013\\
520-560  & 0.15666 $\pm$ 0.00013\\
560-580  & 0.15644 $\pm$ 0.00014\\
580-592 &  0.15638  $\pm$ 0.00027 \\

588.4-590.6 & 0.15703      $\pm$ 0.00011    \\   
592-604 &  0.15631    $\pm$ 0.00022    \\   
604-615 & 0.15617       $\pm$ 0.00036    \\
615-626 & 0.15600     $\pm$   0.00027    \\
626-638 & 0.15610      $\pm$  0.00012    \\

650-700  & 0.15585 $\pm$ 0.00011\\
700-750  & 0.15572 $\pm$ 0.00011\\
750-800  & 0.15586 $\pm$ 0.00012\\
800-850  & 0.15552 $\pm$ 0.00012\\
850-900  & 0.15553 $\pm$ 0.00013\\
900-950  & 0.15546 $\pm$ 0.00013\\
950-1000  & 0.15552 $\pm$ 0.00016\\
1000-1170  & 0.15512 $\pm$ 0.00022\\
1450-1750  & 0.15476 $\pm$ 0.00025\\
1750-2100  & 0.15474 $\pm$ 0.00022\\
2100-2500  & 0.15447 $\pm$ 0.00027\\
3200-3900  & 0.15507 $\pm$ 0.00027\\
4000-5000  & 0.15542 $\pm$ 0.00035\\
5000-6400  & 0.15476 $\pm$ 0.00067\\
6400-9300  & 0.15510 $\pm$ 0.00034\\
23500-24500 & 0.15459 $\pm$ 0.00094\\ \hline
\end{tabular}
\caption{Radius ratio as a function of wavelength, with spot effects accounted for, and binned in 28 wavelength intervals.}
\label{binned}
\end{table}

Shortward of 1 $\mu $m, the broad transmission spectrum is very well measured, and defined by a single steep blueward slope. The core of the sodium doublet is resolved by the higher-resolution STIS data, that also shows an absence of broad wings. The implications were already presented in \citet{pon08}, \citet{lec08}, \citet{sin11} and \citet{hui12}. The overall slope is compatible with Rayleigh scattering by solid or liquid particles (with sizes below 0.1 \micron), with a $\lambda^{-4}$ dependence of the cross section, evenly distributed in an atmosphere at $T\sim 1300$K. The increasing slope towards the UV and the height of the sodium line suggest a temperature rise in the upper atmosphere. The core of the sodium doublet, and possibly the sodium doublet, are visible above the haze Rayleigh signature and constrain the altitude at which the haze is seen: low enough to leave the core of the lines seen, but high enough to cover the pressure-broadened wings of the sodium and potassium doublets. 

The infrared data is compatible with a featureless spectrum, as well as with the presence of  muted molecular features. With the addition of the uncertainties for undetected spot crossings, none of the measured variations in radius in the infrared exceeds the uncertainties. We note that our multi-instrument, multi-transits approach is not best suited to the detection of features in specific spectral ranges. This is better done by analysing data from a single instrument, preferably acquired during a single transit. The issue of infrared features is discussed in detail in \citet{gib11} for the 1-2 \micron\ range and \citet{des11} at longer wavelengths.

It is remarkable, though, that the more data sets accumulate, the
nearer they evolve towards a featureless continuum,
similar to the visible. Indeed, the most economical inference from
the observations would be a monotonic decrease of the transit radius towards longer wavelengths, with no spectral feature rising above the noise level. 

By contrast, clear-atmosphere models struggle to reproduce the observed transmission spectrum of HD 189733b. This is most obvious in the visible, but is also the case in the infrared. Clear-atmosphere models with molecules predict water features between 1 and 2 \micron\ that are incompatible with the NICMOS filter data \citep{sin09}, and a rising opacity and larger transit radius at 8 \micron\  than at 4.5 \micron, that the data do not suggest. An entirely flat spectrum in the infrared produces a comparable fit to the passband-averaged data than the clear-atmosphere model plotted in Figure~\ref{binspectrum} (reduced chi-square near 0.9 in both cases).

\subsection{Simple extended-haze model}

The 0.6-1 \micron\ transmission spectrum from ACS was interpreted by \citet{lec08} and subsequently as the signature of a single high-altitude layer of scattering haze. The grains must be abundant enough to provide a higher opacity than the wings of the sodium and potassium lines, and transparent enough in the visible so that the Rayleigh slope dominates over absorption.  Dust and clouds have been considered as a possible important components in the atmosphere of hot Jupiters since the first atmosphere models and observations. The atmospheric temperature of hot Jupiters like HD 189733b correspond to spectral type L and T for brown dwarfs, and observations have shown that these objects have very red colours \citep{kir00}, interpreted as due to the effect of dust in their atmosphere \citep{cha00}.  Several common grain-forming elements have condensation temperatures above the temperature of late L-type objects, including enstatite (MgSiO$_3$), forsterite (Mg$_2$SiO$_4$), corundum (Al$_2$O$_3$) and elemental iron (Fe). Of these, only enstatite is transparent over the visible wavelength range \citep{dor95,lec08}.

The explanation of the ACS result in terms of a thin, high-altitude layer of enstatite haze made specific predictions about the rest of the spectrum that were not borne out by the observations. The spectrum was expected to flatten towards the UV, as the wavelength becomes comparable to the size of the grains and therefore moves to the flat part of the Mie scattering curve. It was also expected to drop sharply towards the infrared, where the line of sight reaches regions below the haze layer. Measuring these two features was the main motivation of the HST observing programme GO-11740. Neither was corroborated by the new observations with STIS and WFC3. 

The ACS and STIS observations are compatible with the $\lambda^{-4}$ dependence of Rayleigh scattering over more than 5 atmospheric scale heights, i.e. more than two orders of magnitude in pressure  (see Figure \ref{binspectrum}). 
In the infrared the spectrum becomes flatter, and remains above the extension of the Rayleigh slope. 

An extension of the haze hypothesis can account for these observations. Condensates not confined to a high-altitude layer, but extending over most of the atmosphere, can be expected to produce this type of spectrum. At higher altitudes, small grains produce the Rayleigh slope. At lower altitudes, grains become larger, and their size gets closer to the wavelength, producing a flatter spectrum, until a Òcloud deckÓ layer is reached that entirely scatters or absorbs the incoming star light.

A full model of the transmission spectrum of HD 189733b in terms of haze and clouds is beyond the scope of this paper, and probably unwarranted given the present amount of observational constraints. We develop here a first-order, qualitative model to account for the observed spectrum with few assumptions and free parameters.

\subsubsection{Rayleigh slope}

The Rayleigh scattering cross section evolves as:

\begin{equation}
\sigma = \frac{2 \pi^5}{3} \left(\frac{n^2-1}{n^2+2}\right)^2 \frac{a^6}{\lambda^4}
\end{equation}
where $a$ is the grain size and $n$ the refraction index.
Since the dependence on grain size  is very steep , $\sigma \propto a^6$, for most grain size distributions, the largest grains will dominate the scattering cross section. In the case of HD 1897833b, the transmission spectrum implies that the maximum grain size remains similar over five scale heights or more in the upper atmosphere. It also implies that their abundance remains approximately constant. Therefore, keeping the Rayleigh slope in the transmission spectrum over such a large wavelength range requires well-mixed grains with a constant size cutoff.

 \subsubsection{Settling regime}
 
 \label{settling}
The size distribution of grains with altitude for condensates in a planetary atmosphere is controlled by a balance between the timescale for the gravitational settling of the grains, and the timescale for replenishment of the grain population.

At a given altitude, the maximum grain size $ a_{max}$  sustained will be the one for which the two timescales coincide:
\begin{equation}
\tau_{\rm set} (a_{\rm max}) = \tau_{\rm rep} (a_{\rm max})
\end{equation}
where $ \tau_{\rm set}$ and $\tau_{\rm rep}$ are the settling and replenishment timescales.

Following \citet{ack01} and \citet{woi03}, we estimate the settling timescale by calculating the time taken for grains to cross one atmospheric scale height at their terminal fall velocity.The terminal velocity of the particles is obtained  by balancing the pull of gravity against the drag of the gas flow:

\begin{equation}
\tau_{\rm set} = H / v_{\rm fall}
\end{equation}
where $ H = \frac{k T}{\mu m_u g} $ is the atmospheric scale height ($k$ is the Bolzmann constant, $T$ the temperature, $\mu$ the mean molecular weight, $m_u$ the atomic mass unit, $g$ the gravity in the atmosphere).  $v_{\rm fall}$ is the terminal velocity for particles of a given size due to atmospheric drag.

Different expressions for the atmospheric drag can be used depending of the flow regime affecting the grains. For low gas densities, the flow is molecular (Brownian motion). The transition from molecular to viscous flow is described by the Knudsen number, defined as the ratio of the mean free path to the particle size:
\begin{equation}
K_n = \frac{l}{2 a}
\end{equation}

Values  of $K_n $ near 1 separate molecular flow, where individual impacts dominate the dynamics of the particle, from laminar flow, where the gas is dense enough to be treated as a viscous fluid. 

For the grain sizes, gas pressures and temperatures relevant here  ($a <10$ \micron, $p<1$ bar,$ T \geq 1000$ K), the Knudsen number is much larger than unity, so that the flow is closer to molecular than to viscous. In that case \citep{woi03}:

\begin{equation}
v_{\rm fall} = \sqrt{\frac{\pi}{4}} a\,  \frac{\rho_{\rm cond}}{\rho_{\rm gas}}\frac{g}{c_T}
\end{equation}
where $\rho_{\rm cond}$ and $\rho_{\rm gas}$ are the density of the grains and gas, and $c_T$ is the sound speed in the gas:
\begin{equation}
c_t = \sqrt{\frac{\gamma k T}{\mu m_u}}
\end{equation}

Using the ideal gas equation of state, $\rho_{\rm cond}=3.2 \times 10^3$ kg$\,$m$^{-3}$ \citep[][for enstatite]{ack01} and the following parameters for HD 189733b, Ð T=1200 K, g = 21 m$\,$s$^{-2}$, $\mu=2.35$, $\gamma=7/3$,  gives for molecular flow:

\begin{equation}
\tau_{\rm set} =  \frac{p}{[1 bar]}\left( \frac{a}{[1 \micron]} \right)^{-1}  2\cdot 10^8 s
\end{equation}

Thus, for instance, the settling timescale for 0.1 \micron\ at 10 mbar is about eight months.

For the purpose of a first-order understanding of the transmission spectrum, the important parts of the expressions above for $v_{\rm fall}$ is its dependence on particle size and gas pressure:

\begin{equation}
v_{\rm fall} \propto  a^{1} \rho_{\rm gas}^{-1}
\end{equation}

The replenishment timescale $\tau_{\rm rep}$ cannot be estimated robustly from present models and observations. Equilibrium 1-D atmosphere models do not predict the formation of any condensates, because the atmosphere is vertically stable against convection.  3-D models tracking the motion of grains have not been constructed yet for hot Jupiters, although this is expected to change soon.

Two processes could account for grain formation: a vertical exchange of mass in the atmosphere, due to the vigorous day-night atmospheric circulation, for instance via eddy diffusion, or the formation of photochemical haze through the direct action of the stellar irradiation. 
In both cases, the grain formation timescale is not known at present. 
We therefore leave $\tau_{\rm rep}$ as a free parameter. 

In the case of vertical mixing, note that this parameter can be directly linked to the vertical momentum diffusion parameter $K_{zz}$, used for instance in \citet{spi09} and \citet{you10} to parameterise vertical mixing in hot Jupiters:

\begin{equation}
\tau_{\rm rep}  = \frac{H^2}{K_{\rm zz}}
\end{equation}

Values of $K_{\rm zz}$ are found to be $10^3 - 10^5$ cm$^2\,$s$^{-1}$ in brown dwarf models \citep{fre10}. \citet{spi09} find that values up to $10^{11}$ cm$^2\,$s$^{-1}$  are required to keep TiO aloft in hot Jupiters and produce a stratospheric inversion (note that in this context, $K_{\rm zz}$ is used simply to parametrize the amount of vertical mixing, regardless of the actual process responsible for the transport).

A value of $K_{\rm zz}=10^{11}$ corresponds to a replenishment timescale of 1 hour for HD 189733b, a value of $K_{\rm zz}=10^3$ to $10^4$ years.   Note that there is no physical reason for $K_{\rm zz}$ to be constant as a function of height in the atmosphere. On the contrary, large changes are to be expected. Convective mixing for instance tends to produce $K_{\rm zz} \propto \rho^{-1/3}$.

If the formation process produces grains of all sizes, and the size distribution at any given height is dominated by the balance between settling and replenishment, then in the molecular regime, $\tau_{\rm set} \propto a\,p^{-1} $ implies, assuming a constant $\tau_{\rm set}$, 
\begin{equation}
a_{\rm max}(p) \propto p
\end{equation}

This relation between particle size and pressure corresponds to: 
\begin{equation}
d R = -4/7 d \ln(\lambda)
\end{equation}
if the initial grain size distribution is assumed to be flat in terms of number of grains ($n(a) \propto 1$), and 
\begin{equation}
d R = - d \ln(\lambda)
\end{equation}
if the initial grain size distribution is assumed to be flat in terms of mass fraction of grains of difference sizes ($n(a) \propto a^{-1/3}$). The derivation of these relations is given in Appendix C.

\subsubsection{Cloud deck}

A third possible regime is a ``cloud deck'' dominated by large grains. The effect of a layer of clouds with large grains is well approximated by  a linear cut at a given height in the transmission spectrum \citep{sea00}.

\subsubsection{Haze+cloud scenario}

Figure~\ref{grainfit} illustrates these three regimes in the transmission spectrum, (i) Rayleigh scattering by well-mixed small grains, (ii) settling grains, and (iii) cloud deck. It illustrates how a haze+cloud scenario could account for the broad features of the observed spectrum. 

This haze+cloud scenario is partly inspired from examples in the Solar System. Venus and Saturn for instance have an atmosphere dominated by clouds, with a layer of haze above the clouds made of smaller particles, that dominates the opacity to incoming sunlight in the visible \citep[e.g.][]{kno80}.

We note that the observed transmission spectrum suggests a ``bottleneck'' effect on $K_{zz}$. From a certain point upwards, the vertical mixing must grow faster than the settling timescale (thus faster than $p^{-1}$) to keep the maximum grain size constant. In that case, the largest grains that can stay aloft above the cloud deck ($\sim$ 0.1 \micron\ in size) are well mixed all the way to the highest layers. It is the simplest way to produce a constant Rayleigh slope over several scale heights, as observed.

\begin{figure*}
\resizebox{12cm}{!}{\includegraphics{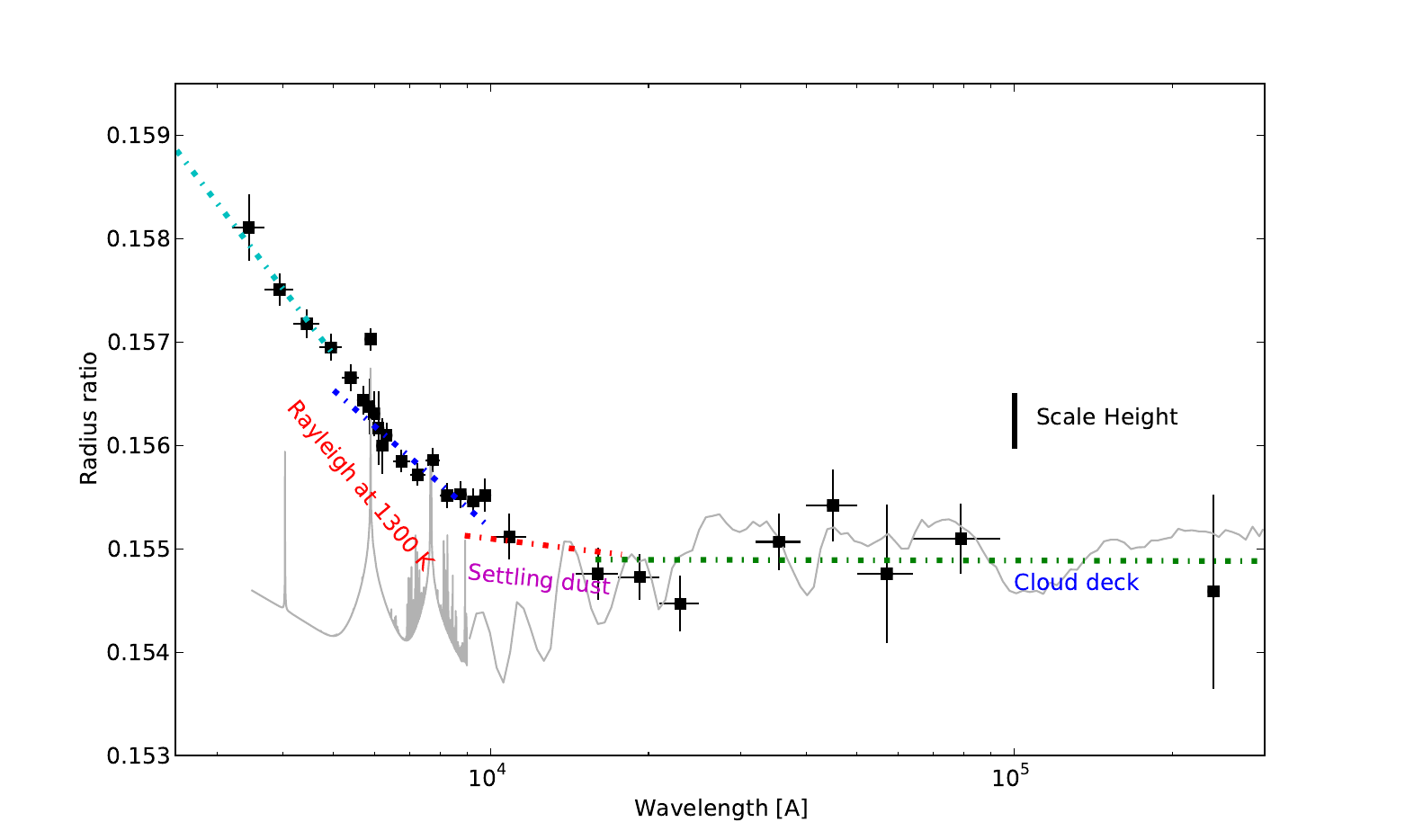}}
\caption{ Spectrum and model as in previous figure. The dotted lines, from left to right, indicate the effect of Rayleigh scattering at 2000K, 1300 K, a cloud with grain sizes increasing linearly with pressure, and an opaque cloud deck. }
\label{grainfit}
\end{figure*}

\section{Haze and clouds in the phase curve and eclipse data}

\label{spec}

\subsection{New/revised eclipse and phase curve data}

Information on the atmosphere of HD 189733b is also available from measurements of the secondary eclipse and phase curve in several infrared channels. The secondary eclipse in a given passband measures the brightness temperature integrated over the whole day hemisphere of the planet. The phase curve measures the day-night contrast in brightness temperature. The phase curve also measure the shift of the hottest spot on the planet, due to the jets redistributing heat from the day side to the night side.

The realisation that the transmission spectrum from UV to mid-infrared is very different from clear-atmosphere predictions leads us to revisit these eclipse and phase curve data as well. It would be surprising that the haze/cloud/dust component were to dominate the transmission spectrum entirely around the whole planetary limb, and have no effect on the emission spectra measured by the eclipse and phase curve. Condensates will not only affect the shape of the emergent spectrum, but by modifying the radiative transfer, affect the redistribution of heat around the planet as well.

\citet[][hereafter K12]{knu12}  presents new eclipse and phase curve data, and a re-discussion of previous data sets, in the Spitzer passbands at 3.6, 4.5, 8 and~24 \micron. Their results are summarised in Table  \ref{knutson2012}.
The seminal results of \citet{knu07} showed that at 8 \micron, the Spitzer light curve indicates an eastward shift of the hot point of $16 \pm 6^o$, and a day-night temperature contrast of $\sim$350~K (K12 show that instrumental systematics make this last number unreliable). The phase curves at 3.6, 4.5 and 24 \micron\ present eastward shifts (listed in Table \ref{knutson2012}) that at compatible with the finding at 8 \micron.

\begin{table}
\begin{tabular}{lllll} \hline
Wavelength&	$\Delta $T &  $T_{\rm day} $& $\phi_{\rm hotspot}$ & $\phi_{\rm hotspot}$  \\ 
\ [\micron]  & [K]	& [K]& observed & model \\ \hline
3.6  &  503 (21$^*$) & 1328 (11)& 35$^o$ (6) & 53$^o$ \\
4.5  &  264 (24$^*$) & 1192 (9)&	20$^o$ (5) & 52$^o$  \\
8 &  350 (-- )  &	 1259 (7) & 23$^o$ (3)    & 49$^o$  \\
24  &  188 (48) & 1202 (46)   & 37$^o$ (8)&  49$^o$  \\ \hline
\end{tabular}
\caption{ Day-night temperature contrast, day-side temperature, and hot-spot eastward longitude shift, observed and expected from a circulation model, for HD 189733b. The temperatures are geometric means over the planetary disc.  Data from \citet{knu12} except the 8 \micron\ and 24 \micron\ day-side temperatures from \citet{cow11}. The uncertainty of the temperature amplitude at 8 \micron\ is unreliable because of instrumental systematics. $(^*)$ according to our re-analysis in Appendix B, the uncertainties of the day-night temperature differences at 3.6 and 4.5~\micron\ should be $\pm$ 150 K and $\pm 78$ K.}
\label{knutson2012}
\end{table}

\subsection{Puzzling features explained by the haze/clouds scenario}

Intriguingly, there are several features of the secondary eclipse and phase curve information that the clear-atmosphere models in K12 struggle to explain, and that would be natural consequences of an atmosphere dominated by haze opacity. 

K12 do not include the condensates in their modelling and discussion of the eclipse and phase curve data. The reason, presumably, is that the emission spectrum probes deeper regions of the atmosphere than the transmission spectrum (by a factor $\sim$ 50, because of the grazing geometry), and that they assume that the haze layer forming the transmission spectrum in the visible is confined to a high altitude and transparent in the infrared. However,  the transmission spectrum data presented in this paper suggests that the haze extends over more than a factor 100 in pressure, and that it may be a source of opacity in the infrared as well.

The first-order effect of dust/haze/clouds is to move the contribution functions and effective photosphere at all wavelengths to higher altitudes (lower pressure). Based on the transmission spectrum, we can expect this to be stronger at shorter wavelengths. Haze does not suppress atmospheric features in the emission spectrum entirely, but can reduce their amplitude, because scattering increases the optical path for a given pressure difference. By moving the photosphere to lower pressures, haze and clouds will also tend to reduce the timescale of heat loss from the atmosphere. In a hot Jupiter, this will tend to decrease the efficiency of the dayside-to-nightside heat redistribution, which can be observed as an increased day-night temperature contrast and decreased eastward drift of the hottest point in the atmosphere.

Four possible signatures of the presence of dust in the eclipse and phase curve information for HD 189733b are listed below. A comparison of the capacity of the ``clear'' and ``dusty'' scenarios to account for the observations is summarised in Table~\ref{compare}.


\subsubsection{ Larger phase curve amplitude at 3.6 microns than in other passbands}

Atmosphere models predict that the amplitude of the day-night temperature contrast increases with altitude, because the radiative timescale increases with higher densities much faster than the speed of the wind decrease \citep[see e.g.][]{sho02,coo05}. In such case, a larger day-night temperature contrast indicates that we see the flux coming from a higher layer in the atmosphere. 

The day-night temperature contrast at 3.6 \micron\ is almost double that measured in other Spitzer passbands (see Table~\ref{knutson2012}). K12 note that this is puzzling for a clear atmosphere dominated by absorption from molecular bands. In any non-pathological configuration, a higher temperature contrast indicates a higher-altitude contribution function. But all models based on molecules abundant in a hot Jupiter atmosphere suggest that the opacity at 4.5~\micron\ should be higher than at~3.6~\micron. The 4.5~\micron\  band should therefore sample a higher layer of the atmosphere, and show a larger temperature contrast. The opposite is observed. 

Equilibrium-chemistry models of hot Jupiters suggest that the opacity at 4.5 \micron\ should be higher than at 3.6 \micron\ \citep[the non-equilibrium models of] [find a similar opacity in the two bands, but significantly underpredict the flux at longer wavelengths measured with Spitzer for HD 189733b]{mos11}.

This is arguably the observation in the \spitzer\ data that is the most robust and tightly connected to the physics of the atmosphere. It is left unexplained in K12. It also affects the interpretation of the day-side emission spectrum, because the assumption of a higher opacity at 4.5 \micron\ is the foundation of the measurement of the temperature inversion in the atmosphere from the \spitzer\ passbands.

However, the presence of condensates can make the atmosphere more opaque at 3.6 \micron\ than at 4.5 \micron\ in spite of molecular absorption. A decrease of opacity with wavelength extending to the near infrared is suggested by the transmission spectrum.

In Section~\ref{discrepant} below and Appendix B, we note that the uncertainties of the phase curve amplitudes at ~3.6 and 4.5~\micron\ may have been underestimated by a large factor in K12, due to an overestimation of the capacity of the gaussian-process interpolation to predict the flux variations on the timescale of the planetary orbit. If our re-assessment is correct, then the difference between the two amplitudes is reduced to the 1-2 $\sigma$\ significance level. Confirming this amplitude difference may be a good case for future Spitzer observations.


\subsubsection{ Reduced hot spot shift}

The longitude of the eastward shift of the hot spot is expected to result from a competition between the heat transfer timescale and the heat loss timescale \citep[identified in hot Jupiters to the advective timescale $\tau_{\rm adv}$ and the radiative timescale $\tau_{\rm rad}$, e.g.][]{sho02}. 

The observed eastward shift of the hot spot in the atmosphere is about half the expected values in the Spitzer channel. K12 discuss several possible explanations: a planet rotating slower than tidal circularisation, enhanced heavy-element abundances in the atmosphere, and an increased atmospheric drag. A slow-rotating planet or increased drag would increase $\tau_{\rm adv}$ by slowing the atmospheric heat redistribution, whereas more heavy elements would reduce $\tau_{\rm rad}$ by allowing more rapid cooling through atomic line emission.

An extended haze cover, however, would naturally bring the hot spot nearer to the sub-stellar point. Increased opacities move the photosphere to lower pressures, where the radiative timescale is shorter. To first order the eastward drift of the hot spot depends on the ratio $\tau_{\rm rad}$/$\tau_{\rm adv}$. $\tau_{\rm rad}$ changes much more rapidly than $\tau_{\rm adv}$ with pressure in hot Jupiters, to first order $\tau_{\rm rad}\propto p^{-1}$ in the upper atmosphere, whereas $\tau_{\rm adv}\propto p^0$. 

Therefore, raising the photosphere by about one scale height will result in decreasing the hot spot shift by $1/e$. This is compatible with the indications from the transmission spectrum. The infrared opacities are roughly one scale height higher in the transmission spectrum than the clear-atmosphere expectations, which would predict a  latitude for the hot spot reduced by the same factor compared to the clear-atmosphere models, as suggested by the observations.

\subsubsection{ Large brightness temperatures at 8 \micron\ and 24 \micron} 

K12 conclude that the observed 8 \micron\ and 24 \micron\ brightness temperatures are larger than the values predicted by clear-atmosphere models, both on the day side and on the night side. They are nearer to a black-body curve than to the expected spectrum sculpted by molecular absorption lines (see Fig. \ref{K12bb}).
K12 find the observations impossible to fit even with the added free parameters of a high-altitude grey absorber and non-equilibrium chemistry. Only modifying the planet rotation rate and the elemental abundances produces a satisfactory fit.

However, a brightness temperature at 8 and 24 \micron\ closer to the shorter wavelengths than for a clear-atmosphere models is a natural prediction of a Òhaze and cloud-deckÓ configuration. Scattering and absorption by condensates shorten the altitude differences sampled by the different wavelengths. For a given temperature gradient with altitude, the brightness temperatures will be correspondingly closer, bringing the emergent spectrum nearer to a blackbody.

\subsubsection{ Flat spectrum at 5-14 \micron}

The observations at 3.6, 4.5, 5.8, 8, 16 and 24 \micron\ with \Spitzer\ are passband-integrated time series, and therefore do not provide information on the local shape of the emissions spectrum. \citet{gri07} observed the eclipse spectrum between 5 and 14 \micron\ with the IRS spectrograph on \Spitzer. These results show a less marked feature than in clear-atmosphere models  (see figure~10 of K12, repeated in Fig. \ref{K12bb}). Molecular features are predicted to cause a drop in the flux ratio beyond 10 \micron\ that is not observed in data. In best-fit model from K12, for instance, the uncertainties must be stretched to make the observed spectrum marginally compatible with the model. While this configuration produces a satisfactory $\chi^2$ statistics, it is improbable: random errors do not usually conspire to erase spectral features.

Again, lower-amplitude features are a natural prediction of cloudy/haze models. Reducing the amplitude of molecular feature is an observed effect of condensates in L-type brown dwarf spectra.

\subsection{``Dusty'' scenario for HD 189733b}

Altogether, the simplest scenario compatible with the transmission spectrum -- i.e. a gentle overall decrease of the opacity with wavelength -- also explains observed anomalies in the emission and phase curve data.  A global haze/cloud, with grains large enough near the infrared photosphere to affect radiative transfer,  moves the photosphere upwards, reduces the hot spot shift, and damps the differences between the depth of the contribution functions for the different Spitzer channels. It may also make the atmosphere more opaque at 3.6 \micron\ than at longer wavelengths.

K12 explain the above features in a piecemeal fashion, invoking several additional ingredients, such as enhanced metallicity, non-equilibrium chemistry, absence of tidal synchronisation. All these explanations are plausible.  Still, none of the clear-atmosphere models can explain a high day-night contrast at 3.6 \micron.

The introduction of  ``dusty'' models was a leap forward in the study of brown dwarfs \citep[see e.g.][]{all01,kir05,sau08}, and our results suggests that a similar effort may be useful for hot planets like HD~189733b.


\begin{table*}
{\large
\begin{tabular}{lcc}

Observation   &  Dusty atmosphere & Clear atmosphere \\ \hline

Shortwave transmission spectrum & ok & no \\

Higher day-night contrast at 3.6 \micron & ok? & no \\

Longwave day-side brightness  & ok & no \\

Featureless IRS spectrum & ok & difficult \\

Low hot-spot shift & ok & difficult \\

High transmission opacity at 4.5 \micron & difficult & ok?* \\ \hline

\end{tabular}
}
\caption{Compared capacity of ``dusty'' and ``clear'' atmosphere scenarios to account for  observations. *: requires non-equilibrium CO absorption according to K12.}

\label{compare}

\end{table*}

\subsection{Remaining discrepancies}

\label{discrepant}

Some features of the observations do not fit neatly in the simplest models, even when including the possible effects of condensates. 

One is the marginal indication in the transmission spectrum that the radius at 4.5 \micron\ is larger than at 3.6 \micron. Uncertainties due to the effect of star spots on the transit depth (discussed in Section \ref{starspots}) imply that this is only at the $\sim 1-\sigma$ significance level, although both \citet{des11} and K12 observe a larger radius at 4.5 \micron.

This would be difficult to reconcile with the phase curve amplitude being larger at 3.6 \micron. Only large changes in the relative opacities between the two channels around the planet could account for both features.

There is also a marginal indication of a lower opacity at 3.6 \micron\ from the higher hot spot shift. The simplest haze model would predict the phase shift to decrease with increased opacity. This is a less stringent constraint than the day-night temperature contrast though, because the longitude of the hot spot depends on the details of the atmospheric circulation, and the $\tau_{\rm adv}/\tau_{\rm rad}$ approximation is rather crude.

We note that there is a tension between the phase curve, hot point shift and transit depth results at 3.6 and 4.5 \micron in K12, in both the clear and dusty scenario. Given the importance of star spots and instrumental effects, it is possible that one of these results will turn out to be modified in light of further observations.

K12 give the highest significance to the phase curve amplitude,  and consequently this observation has a strong weight in our interpretation.  However,  the phase curve of the planet is entangled with the variability of the hot star in the infrared light curves. K12 use the Gaussian-process interpolation of the APT light curve given in Appendix~A to constrain the variation in the stellar flux during the planetary orbit to recover the phase curve. There are two pitfalls with this procedure though. The first is that there is no APT measurements close in time to the \spitzer\ measurements, so that a large extrapolation over time is required. The second is that the time sampling of the APT light curve (one point per day typically) is not adapted to resolving the curvature of the light curve on the time scale of the planetary orbit (2.2 days). This implies that the constraint on the shape of the stellar contamination to the measured phase curve will depend mainly on the assumptions made on the covariance kernel of the Gaussian process, rather than on the data itself.

In Appendix B, we study how these two factors may have led to an under-estimation of the uncertainties on the phase curve amplitudes  at 3.6 and 4.5 \micron\ in K12. We are helped by the availability of one month of precise continuous monitoring of HD 189733b with the MOST satellite. This provides an empirical check on the capacity of the Gaussian-process interpolation to constrain the curvature of the stellar light curve over the time scale of the planetary orbit. 

We find that indeed, the capacity of the Gaussian-process interpolation to constrain the stellar contribution  is severely limited. Consequently, we recalculate the uncertainties on the phase curve amplitude as $27 \cdot 10^{-5}$ and $25 \cdot 10^{-5}$ for 3.6 and 4.5 \micron\ respectively (instead of $6 \cdot 10^{-5}$ and $9 \cdot 10^{-5}$ in K12). This translates in uncertainties of $\pm$150~K and $\pm$78~K on the temperature contrasts (instead of $\pm$21~K and $\pm$24~K). The details are given in Appendix B. With these uncertainties, the phase curve amplitude at 3.6 \micron\ is still larger than at 4.5 \micron, but with a significance reduced to $\sim 2 \sigma$.

\section{A new picture of the atmosphere of HD189733b}

In this Section we explore some consequences of the haze+cloud scenario for the planet HD189733b.

\label{toy}

\subsection{Is a hot stratosphere ruled out?}

\label{hotstrat}

The presence of an inverted temperature profile near the photosphere has been inferred for several hot Jupiters from the relative fluxes in the 3.6 and 4.5 \micron\ \spitzer\ bands  \citep{for10}. Since the opacity at 4.5 \micron\ is assumed to be higher because of molecular bands, a higher surface temperature at 4.5 \micron\ is interpreted as a sign of a temperature increasing with height in the atmosphere. Some hot Jupiters are thus thought to possess ``hot stratospheres''. likely caused by a visible-light absorber at high altitude. 

Since the brightness temperature observed for HD~189733b at 4.5~\micron\ is smaller than at 3.6~\micron, a normal temperature profile (i.e. decreasing with height) is inferred.

However, the ``dusty'' scenario puts this argument on its head. If the opacity difference between the two bands is inverted, as suggested by the phase curve results, then the relation between the 3.6/4.5 temperature difference and the temperature profile is inverted as well. A smaller brightness temperature at 4.5 \micron\ now implies a temperature profile {\it increasing} with height, i.e. a ``hot stratosphere''. In this interpretation, HD 189733b would have an inverted temperature profile as well. The relation between the temperature brightness at 3.6 and 4.5 \micron\ would be caused by an inverted ratio of opacities, combined with a rising temperature with altitude.

\begin{figure}
\resizebox{8cm}{!}{\includegraphics{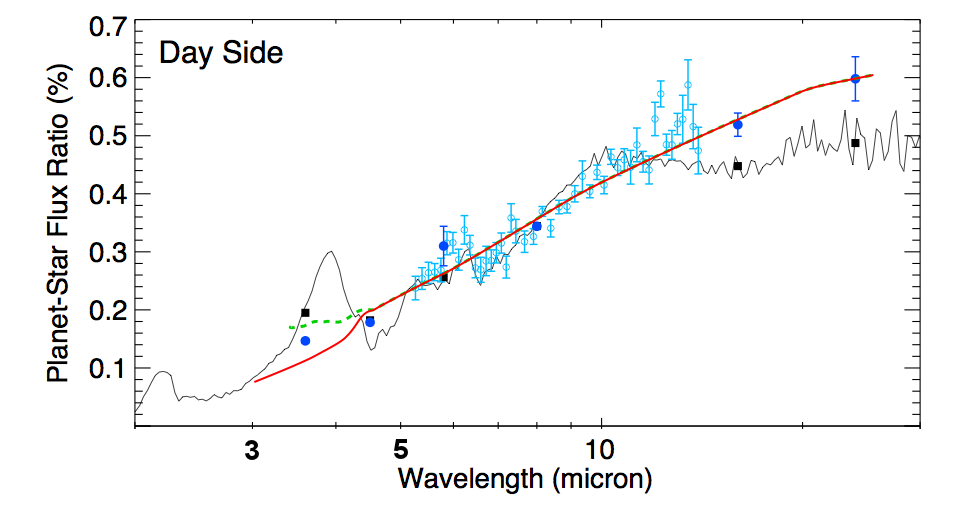}}
\caption{\Spitzer\ data for the day-side emission spectrum of HD 189733b (blue) and best-fit solar-abundance spectrum from K12 (black) -- adapted from figure 10 in K12. The red line shows a blackbody spectrum at 1200~K for the atmosphere of the planet (assuming 5000~K for the star). The dotted green line shows the expectation from an empirical model based on the transit and phase curve data.}
\label{K12bb}
\end{figure}

We note that an inverted temperature profile is a natural consequence of the ``dusty'' picture. If the opacity decreases from visible to near-infrared, it implies that the incoming flux from the star is absorbed higher than the outgoing thermal flux is emitted, i.e. an ``anti-greenhouse'' situation. This tends to produce a temperature inversion above the infrared photosphere (like the ozone layer on Earth).

In the transmission spectrum,  the opacity in the visible is clearly higher than in the infrared, by about an order of magnitude. If that extends to deeper layers, the anti-greenhouse effect will be very strong.
 
To first order, the sign of the T-P profile near the photosphere for a hot Jupiter is set by the ratio of opacities at the wavelengths of the incident starlight to the opacity at the wavelengths of the outgoing thermal radiation from the planet. In the notation of \citet{hub01}, a ``Greenhouse factor'' $\gamma$ can be defined as $
\gamma = \kappa_{\rm s}/\kappa_{\rm l} $
where $\kappa$ are the flux-weighted opacities and $s$ and $l$ denote the shortwave  and longwave opacities.
$\gamma\!>\!1$ corresponds to inverted temperature profiles.

This picture is modified, however, if the dust affects the opacities by a mixture of scattering and absorption. \citet{hen12} considered the effect of scattering in the simplified analytical framework of \citet{gui10}. Then the temperature-pressure profile will depend on the absorption to scattering ratio.

We computed the expected T-P profile from the \citet{hen12} relation, assuming that the haze is purely scattering and that the absorption is dominated by the atomic and molecular lines expected from the models. This corresponds to $\gamma = 10$ and $\xi = 0.1$ in the notation of \citet{hen12}. $\xi$ parameterises the ratio of scattering to absorption in the shortwave.
Figure \ref{TP} shows the resulting T-P profile. A gradient of $d T / d (z/H) \sim 200$ K is found between the longwave and shortwave photospheres (we note that the analytic approximation predicts an isothermal profile near the infrared photosphere, but this is a probably unrealistic feature due to the two-band approximation used in separating the flux into a shortwave and longwave component).

\begin{figure}
\resizebox{8cm}{!}{\includegraphics{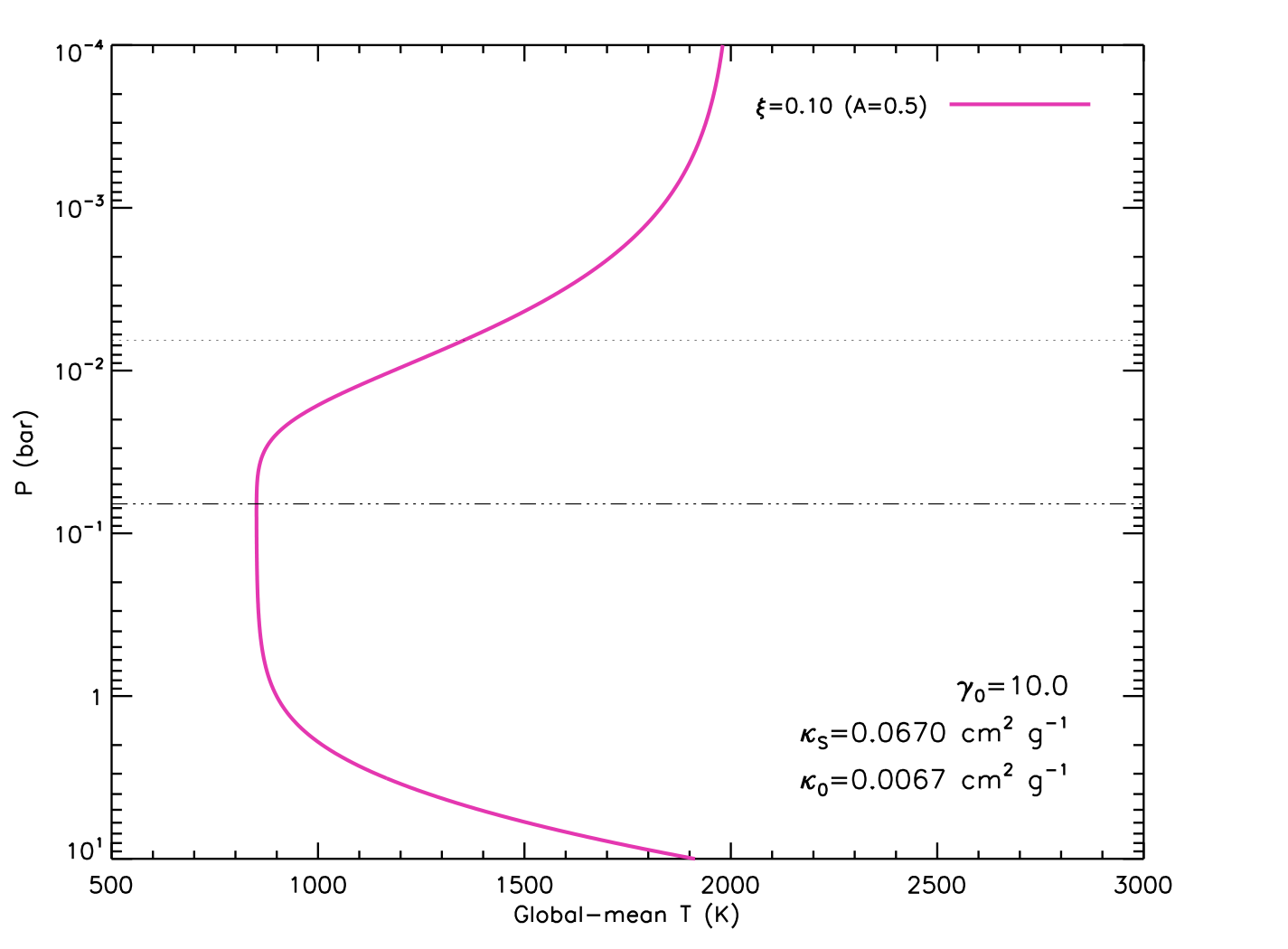}}
\caption{Temperature-pressure profile from the Heng et al. (2011) analytic model for $\gamma_0 = 10$ and $\xi = 0.10$, corresponding to a strong anti-greenhouse effect by a highly scattering dust. The resulting visible albedo is 0.5, $\kappa_s$ and $\kappa_0$ are the shortwave and total absorption coefficients. The dotted and dash-dotted lines indicate the position of the visible and infrared photospheres.}
\label{TP}
\end{figure}

This value is compatible with the measured dayside temperature difference between 3.6 and 4.5 \micron\ (136 $\pm$ 14 K) combined with the pressure difference inferred from the phase curve amplitude ($\sim 2$), which corresponds to $dT / d(z/H) \sim 136/(2/e) = 184$ K.

The green line on Fig. \ref{K12bb} shows the expected secondary eclipse spectrum obtained from a model inspired entirely from the observations, without assumptions from atmosphere models. The opacity is assumed to be constant with wavelength throughout the infrared, following the transmission spectrum, except in the 3.6~\micron\ channel. At 3.6~\micron\ the photosphere is assumed to be one scale height higher in pressure, as suggested by the observed day-night temperature contrast. The temperature gradient near the photosphere is taken to be +200 K per scale height, as given by the analytic expression with a scattering haze. We adopted a T=5000 K Kurucz spectrum for the star. 

The result is remarkably close to the observed \spitzer\ data, given the lack of tuneable parameters.

\subsection{Pervasive dust}

We therefore converge on the following model for the ``dusty'' scenario of the atmosphere of HD 189733b:

-- Grains are present throughout the transparent part of the atmosphere. The mean size of grains decreases towards lower pressures, because of the balance between uplifting by vertical mixing, and gravitational settling. 

-- Scattering by dust dominates the transmission spectrum, causing the transit radius to diminish monotonically from 300 nm to 1 \micron, and possibly all the way to the mid-infrared, masking the pressure-broadened wings of alkali-metal absorption and most molecular bands.

-- Dust scattering and absorption in the visible, above the level of the thermal-infrared photosphere, causes a temperature inversion. The corresponding anti-greenhouse effect cools the mid atmosphere.

-- The opacity decreases significantly from 3.6 to 4.5 \micron\ because of large dust grains, causing the phase curve to be more pronounced in the first passband than in the second, contrary to expectations if molecules dominate the infrared opacities. 

-- Because the heat is deposited and transported at lower pressure than in the clear-atmosphere case, the radiative timescales are lower, and the eastward shift of the hot-spot consequently lower. Another observable consequence of the same phenomenon is that the day-night amplitude is higher than expected from clear-atmosphere models, particularly at 3.6 \micron.

Such highly modified opacities compared to a dust-free model, especially in the visible, will have significant consequences for the atmospheric circulation.  Overall, the deposition of incoming starlight will be moved to lower pressures, lowering the radiative timescale. This will tend to make the redistribution of heat by atmospheric winds less efficient.  \citet{per12} have studied the first order effect of varying the ratio of visible to infrared opacities on the circulation and the eastward drift of the longitude of the hot spot. 

The higher deposition of stellar energy in the atmosphere will also affect the energy budget of the planet. Many hot Jupiters have anomalously large radii, and this is thought to be due to a transformation of some of the incoming stellar radiation energy into internal entropy in the planet. The exact mechanism for this transformation is not yet determined. The two leading candidates are (i) deep dissipation of the day-night currents \citep{sho02} (ii) magnetic drag \citep{bat10,per10}. If the deposition of incoming starlight is moved to lower pressure, the efficiency of both mechanisms could drop considerably.

Obviously, the ``dusty'' interpretation of HD 189733b is at this stage only a possibility. The issue is under-constrained by the present data. Given the complexity of planetary atmospheres, the patchiness of the observations, and the difficulty of controlling instrumental effects in the data, a clear-atmosphere interpretation with various additional effects remains a valid option. 

What kind of observations could allow us to discriminate between the two pictures? One prime candidate is the observation of the reflection spectrum of the planet in the visible and near-infrared. This is challenging but within reach of HST. A high albedo in the visible would indicate that scattering by condensates affects the zenith geometry as well as the grazing incidence. On the contrary, a very dark albedo would suggest that the hazes seen in the transmission spectrum are restricted to a high altitude and do not affect the visible opacities near the photosphere. The colour dependence of the albedo would provide a clue on the importance and grain size distribution of the clouds. \citet{ber08,ber11} have claimed the detection of a high albedo in polarized light for HD 189733b, however these results were not confirmed \citep{wik09}.

Given the large amount of observations devoted to this planet, and the difficulty of correcting for the variability of the parent star, another way forward is to collect similar observations for other planets, in the hope that, as happened for brown dwarfs, ensemble data will be suggestive of the overall effect of condensates if they are a dominant factor for some hot Jupiters. Ongoing HST and Spitzer observation campaigns will address this issue. A dozen hot Jupiters may soon be observed with enough accuracy for a tentative examination of the ``dusty atmosphere'' hypothesis.

\subsection{Origin and composition of the condensates}

Thermal condensates are the most obvious candidate for the haze and clouds of HD 189733b. Scattering needs to dominate over absorption in the visible for the small, high-altitude grain, which favours $\leq 0.1$ \micron\ enstatite grains. A solar abundance of silicates provides enough grains to account for the observed height of the scattering \citep[see][]{lec08}.

Since the condensation temperature of silicates is above 1500 K and the photosphere temperatures in HD189733b are in the 900-1400 K range, these grains would have to form in deeper parts of the planetary envelope, then be transported upwards. This is more difficult to conceive in hot Jupiters than in brown dwarfs, because the strong stellar forcing suppresses convection in the atmosphere. However, the day-night recirculation may imply a substantial amount of vertical motion, that can mix condensates where they would not form under equilibrium condition \citep{par13}. In some models, convection operates in the cooler night side \citep{bur05,dob08}. Gas-phase silicates dredged up from the night side could spread to the whole atmosphere, in equilibrium with gravitational settling.

One drawback of this scenario is that silicates are strongly absorbent in the mid infrared (beyond 8 \micron), but the transmission data does not suggest increased absorption in the 8 \micron\ and 24 \micron\ bands.

Another type of candidate for condensates in hot Jupiter atmospheres are photochemical compounds, for instance sulphur and carbon ÔsootsÕ produces by stellar UV radiation in the upper part of the atmosphere. Because HD 189733 is a very active star, the planet receives a large amount of UV, making photochemistry more likely. According to \citet{zah09b,zah09} though, carbon photochemical products are expected to be highly absorbent, and sulphur products should produce a large feature shortwards of 400~nm, neither of which correspond to the observed transmission spectrum of HD 189733b. 

These two possibilities are not exclusive. Photochemical hazes can seed the formation of condensate clouds. This kind of interaction is known to operate on Earth and in planetary atmospheres in the Solar System.



\subsection{Re-interpreting the two classes of hot Jupiter atmospheres}

There is a stark contrast between the two hot Jupiters for which extensive spectroscopic data has been obtained, HD~189733b and HD~209458b. The atmosphere of HD~209458 b appears very transparent, with a low albedo, and sodium and possibly titanuim oxide absorption in the red, and so transparent in the blue that scattering by the Hydrogen molecule becomes detectable. By contrast, the haze on HD~189733 b is sufficient to elevate the effective transit radius to a few millibars. 

HD 209458b is the prototype of the Òinverted temperature profileÓ hot Jupiters (measured by $T_{3.6}<T_{4.5}$), while HD 189733b represents the class with  $T_{3.6}>T_{4.5}$. \citet{for08} used this dichotomy to separate hot Jupiters in two classes, the "pM" and "pL" classes, in analogy with the M and L spectral types of brown dwarfs. They speculate that in the hotter pM atmosphere, gaseous TiO and VO absorbs the visible star light above the infrared photosphere, thus causing a temperature inversion, whereas in the pL classes, the temperature is low enough for TiO and ViO to condensate out of the gas phase.

Subsequent data have not supported a clean correlation between atmospheric temperature and the $T_{3.6}/T_{4.5}$ ratio.  \citet{knu10} has noted a strong correlation between activity in the host star and apparent temperature inversion inferred from the emission spectrum, which could correlate the temperature inversion with the UV flux of the star instead.

However, according to our results, it is possible that both HD 189733b and HD 209458b have inverted temperature profiles, the difference being that dust on HD 189733b modifies the opacities so that the effective temperatures in the 3.6 and 4.5 \micron\ bands are swapped (see paragraph \ref{hotstrat}). Could this dichotomy be representative of two important classes of hot Jupiters? In that case, the two categories of hot Jupiters in $T_{3.6}/T_{4.5}$ ratio would not be caused by a {\it temperature} inversion, but by an {\it opacity} inversion between the two bands, relative to clear-atmosphere models.

HD 209458b and HD 189733b are representative of their class in terms of stellar activity as well as temperature. The host star HD 209458b is very quiet, while HD 189733 is very active. Thus we need spectroscopic data for more object to understand if the presence of condensates responds to temperature or to UV irradiation. A dependence with temperature would suggest condensation clouds, whereas a link with UV activity would suggest photochemical processes. 

The two categories of hot Jupiters would then correspond to clear atmospheres around quiet stars, and ÔdustyÕ atmosphere around active stars. This hypothesis can be tested on present and future observations of hot gas giant planets with \Spitzer\ and HST.

\subsection{Link with brown dwarfs and young Jupiters}

The temperatures in the atmospheres of hot Jupiters is comparable to that of L-T type brown dwarfs. It is thought that the colours of L dwarfs are explained by the appearance of clouds, which then sink below the photosphere and become invisible in T dwarfs \citep{bur06,sau08,ste09}. 

Brown dwarf atmospheres have two fundamental differences compared to hot Jupiters: the gravity is much higher, and the dominant source of energy transfer is convection rather than radiation and advection. Nevertheless, the similarities in temperature and composition are sufficient for brown dwarf studies to inform our understanding of gas giant planets \citep{cur11}. 

HD 189733b has a temperature that falls in the T-type range (700-1400 K). However, \citet{mar12} pointed out that, from the point of view of clouds, planets and brown dwarf temperatures should not be compared directly.  Other things being equal, the effect of a smaller gravity allows for the persistence of clouds above the photosphere at lower temperature. The fundamental reason is that, at lower gravity, a given pressure corresponds to more mass. Because the opacity of grains is independent of gas pressure, while the opacity of atomic and molecular lines grows with pressure, cloud opacity will remain important longer at lower gravity. 

Figure \ref{Tg} shows the position of some brown dwarfs,  hot Jupiters and directly imaged planets in a temprature-gravity diagram. The dashed line shows the gravity dependence of the disappearance of clouds below the photosphere according to \citet{mar12}. Grey zones and grey symbols indicate the possibly ``dusty'' objects, i.e. atmospheres where haze/cloud opacities seem required to explain the observed spectra.
The atmosphere of the wide-orbit, young  planets found by direct imaging around HD 8799 \citep[e.g.][]{mar08}  have temperatures similar to HD 189733b. 
The presence of clouds has been inferred in their atmosphere from their infrared colours \citep{cur11}. 

This provides another angle to tackle the behaviour of clouds in the gravity-irradiation plane of sub-stellar objects. In spite of the fundamental difference in irradiation regime and gravity, the observations suggest the default hypothesis that the same kind of clouds form in hot Jupiters than in less irradiated young planets of the same temperature, and hotter brown dwarfs.  

As more planets and brown dwarfs are characterised by direct-imaging observations, and more spectroscopic data are collected on hot Jupiters and cooler transiting Jupiters, we may hope to obtain a fuller picture of this Òhot substellarÓ class of atmospheres and the role of high-temperature clouds.

\begin{figure*}
\resizebox{16cm}{!}{\includegraphics{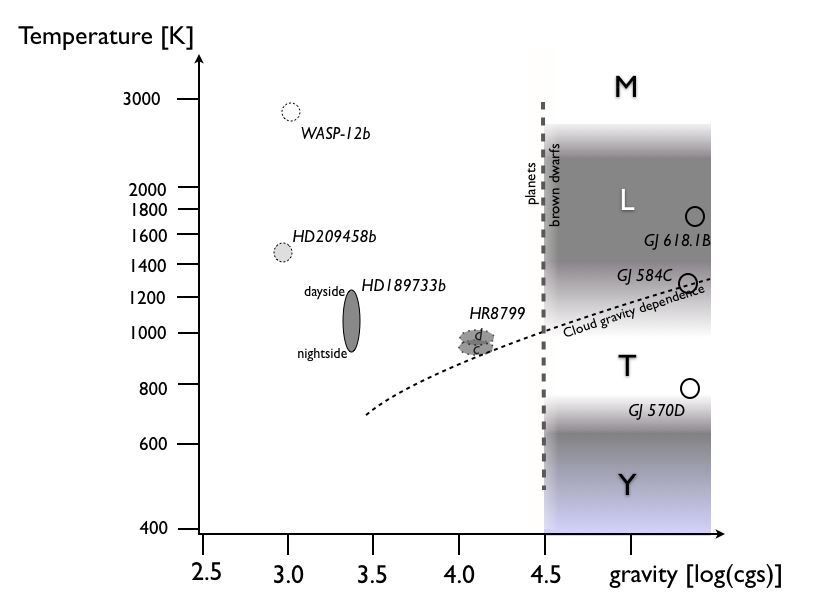}}
\caption{Brown dwarfs and planets in a temperature-gravity diagram, showing the approximate position in temperature of the four classes of brown dwarf spectra, M, L, T and Y. Grey areas show the domains where the infrared colours are believed to indicate the presence of clouds, i.e. L type and possibly Y type. Ellipses show the position of the direct-imaging planets HD 8799c,d, HD 189733b, and two other well-studied hot Jupiters,  HD209458b and WASP-12b. Grey ellipses indicate objects possibly dominated by clouds. The gravity dependence of cloud effects on brown dwarf and planets is from \citet{mar12}. The dotted line shows the $M=13 M_J$ planet/brown dwarf limit for $R =1 R_J$. {\small \it Data for HR 8799 from \citet{mar12}, for HD 189733b from \citet{knu12}, for HD 209458b and WASP-12b from \citet{cow11}.}}
\label{Tg}
\end{figure*}

\section{Summary and conclusion}

In this paper we have attempted to build a complete transmission spectrum of the atmosphere of the hot Jupiter HD~189733b from the UV to the infrared. We have used three different instrument on \HST\ to cover the range from 300~nm to 1~\micron. Other \HST\ and \Spitzer\ observations extend the coverage to 24 \micron\ in a sparser manner.   

The presence of star spots on the surface of the  host star HD 189733, and of instrumental systematics in the space data at the high level of precision required, dominate the error budget. We have devoted much care to accounting for these two factors. In both cases, we have used Gaussian processes to model the nuisance factors in a non-parametric and Bayesian way. These approach yields larger but, we believe, more realistic uncertainties than the usual parametric fits to both the instrumental systematics and the star spot variability.

We find that the atmospheric transmission spectrum of HD 189733b is globally featureless from 300 nm to 1 \micron, rising towards the blue with a slope compatible with Rayleigh scattering by small ($<0.1 $\micron) grains of condensates. The strong cores of the sodium and (possibly) potassium doublets are the only features rising above the continuum.

In the infrared, 1 - 24 \micron, we find that the uncertainties preclude any definite conclusion about the shape of the transmission spectrum. The data is compatible with an extension of the featureless spectrum in the visible, with a flattening of the slope, or with weak molecular features. Individual datasets not provide evidence for the clear-atmosphere, solar-abundance models.

Overall the transmission spectrum suggests an extended presence of haze/clouds in the atmosphere of the planet.

We then combine the information from the transmission spectrum with the indications from the day-side emission spectrum and the phase curves measured with \Spitzer\ in the infrared. We find that several of the anomalies in these data could be explained by the prevalence of condensates in the atmosphere. Notably, the high amplitude of the phase curve at 3.6 \micron\ is difficult to reconcile with clear-atmosphere models, and emerges more naturally in the presence of condensates. The lower-than-expected eastward shift of the hottest point on the day side would also be a natural consequence of the presence of clouds.

Overall, while several interpretations remain possible, we find that the present data suggest the possibility that opacity from condensates dominates the atmosphere of HD 189733b, with important consequence not only on the transmission and emission spectrum, but also on the atmospheric structure, circulation and evolution. In particular, the dichotomy between hot Jupiters with "hot stratospheres" (i.e. temperature inversion near the photosphere) and without, may in that case be instead a dichotomy of hot Jupiters with and without dust, both classes having a hot stratosphere.

Placing HD 189733b in the wider context of gas giant planets and brown dwarfs, we suggest two possibilities: (1) the two classes of hot Jupiters corresponds to clear-atmosphere planets and to planet in which a high UV flux from the active parent star trigger the formation of a photochemical haze and clouds (2) hot Jupiters like HD 189733b have the same silicate/iron clouds as L-type brown dwarfs, and as young planets found by direct imaging like the companions of HR 8799. 

Both hypothesis are testable in the near future by new observations. It is necessary to stress that the interpretation of the data on HD 189733b presented here is tentative, and is not unique. Observational uncertainties are large and sometimes poorly constrained, some observations are only marginally compatible with others, and the models have many parameters since they have to account for global features of the whole planet. A large amount of space-telescope time has been devoted to HD 189733b, and it is probable that definite progress will have to wait for a new generation of observatories, such as JWST or dedicated space projects. In the meantime, more insight may come from the characterisation of other comparable planets. A few more hot Jupiters are accessible to spectroscopic characterisation with \HST, and new young planets are expected to be discovered in the coming years by direct-imaging programmes.

Our results also have implications for the study of exoplanet spectra, and the search for biomarkers on terrestrial exoplanets in the longer terms.  By necessity, information is often inferred for exoplanets by fitting suites of model spectra in a Bayesian fashion to a few Spitzer passbands, sometimes complemented by a few ground-based measurements at shorter wavelength \citep{mad11,lee12}. These ``spectral retrieval'' suites generally do not include condensates and clouds (which would make the Bayesian integration intractable and too dependent on prior assumptions about the properties of the clouds). If our experience with HD 189733b is any guide, we caution against taking these results too seriously.  Those results may be invalidated by the contribution of haze or clouds. The implication of the case of HD 189733b could be summed up as:  Òbeware of incomplete spectraÓ. It is clear that fitting a suite of synthetic spectra to a restricted subset of the data considered in this paper would lead to very misleading conclusions. Each time more extended data has been forthcoming, it has flatly defeated the expectation. 

This has implications for the design of instruments and space missions for the study of exoplanet atmospheres. It would tilt the balance towards an extended spectral coverage, not too narrowly focussed on expected features and model predictions, and keeping the possible presence of condensates in mind when calculating detection capabilities. This may also extend to the search for biomarkers in habitable exoplanets.

Our results suggest several lines of investigation for hot Jupiter atmosphere models. 3-D circulation models can trace the formation and circulation of dust in a hot Jupiter like HD 189733b, and test whether the distribution implied by the transmission spectrum can result from silicate grains. Atmosphere structure models can add haze/clouds opacities to the radiative transfer terms and study the consequences on the observable signatures, such as phase curve,  photosphere temperature and wind speed. Models of photochemical pathways can explore the possibility that small, transparent grains can be produced by stellar UVs in the upper atmosphere of hot Jupiters.

\section*{Acknowledgments}

The authors acknowledges support from the Science and Technology Facilities Council (STFC grants ST/F011083/1 and ST/G002266/2) and the Halliday Foundattion, GWH acknowledges support from NASA, NSF, Tennessee State University, and the State of Tennessee through its Centers of Excellence program. This work is based on observations with the NASA/ESA Hubble Space Telescope, obtained at the STScI operated by AURA, Inc., and observations made with the Spitzer Space Telescope, which is operated by the Jet Propulsion Laboratory, California Institute of Technology, under contract to NASA.

We are indebted to the anonymous referee for a very thorough and thoughtful report. We thank David Charbonneau, Jean-Michel Desert, Ron Gilliland, Heather Knutson and Alain Lecavelier for their contributions to the GO-11740 HST proposal, Stephen Roberts and Mike Osborne for guidance on GP regression, and Christiane Helling, Kevin Heng and Jonathan Fortney for insightful comments on the first version of the manuscript.

\bibliographystyle{mn2e_astronat}
\bibliography{alltogether}{}

\clearpage

\appendix

\section[]{Gaussian process interpolation of the stellar flux}

In this Appendix, we explain how we used ground-based observations of the brightness of HD\,189773, collected over several years, to evaluate the impact of unocculted star-spots on transmission spectra obtained during the same period with HST and \Spitzer. Our goal is to estimate the fraction of the stellar disk that is covered in spots at the time of each transit observation, and the resulting dimming of the star (relative to the unspotted flux, which is not known) in the relevant bandpass.

We proceed as follows:
\begin{enumerate}
\vspace{-1mm}
\item combine long-term out-of-transit monitoring of HD\,189733 from several observatories in different bandpasses;
\item model the resulting time-series using a quasi-periodic Gaussian process (GP) model;
\item estimate the star's brightness at the time of each transit observation, relative to a putative un-spotted brightness;
\item convert this to the bandpass of each observation;
\item use the results to convert the measured planet-to-star radius ratio in each band.
\end{enumerate}
\noindent There are 2 free parameters in this process: the spot temperature, used in steps (i) and (iv), and the un-spotted flux, used in step (iii).

\subsection*{Combining datasets}

Our main source of out-of-transit data for HD\,189733 is the long-term monitoring program carried out with the 0.8-m Automated Patrol Telescope (APT) at Fairborn Observatory \citep{hen08}. The APT uses photomultiplier tubes to gather photometry in the Str\"omgren $b$ and $y$ filters simultaneously, resulting in an effective `$b+y$' bandpass. This program started in 2005 and is ongoing, with two-month-long breaks each semester (in August and January). There are typically one or two points per night, resulting in a total of 868 observations.

We also used observations taken with the 40-inch telescope at Wise Observatory in November 2009 using the Bessel $R$ filter \citep{sin11}. There were 24 observations, with a typical sampling of 1 point per night. To account for the difference in bandpass with the APT data, the Wise data were scaled by a factor $1.12$. This factor is simply $R^{\rm mag}_{ij}(\delta, T)$, as defined in Section 2.2.4, where $\delta= 0.01$, $i$ and $j$ corresponding to the Wise $R$ and APT $b+y$ bandpasses respectively, and we adopted $T=4250$\,K for the spots.
Furthermore, a constant offset was added to the Wise data, to bring its median magnitude level with that of the APT data in the two nearest semesters.

\subsection*{GP regression basics}

We model the out-of-transit brightness variations of HD\,189733 as a GP, which enables us to interpolate these measurements to the times of the HST and \Spitzer\ observations, subject to certain smoothness constraints. Specifically, we use a quasi-periodic kernel to account for the fact that the spot-induced variability has a dominant periodicity (the rotation period of the star), but its amplitude and phase evolve. An important feature of the GP model is that it predicts a probability distribution for the star's magnitude at any specific date, i.e.\ our interpolates are associated with robust error bars.

Here we summarise the main steps in GP regression very succinctly. The interested reader will find an excellent textbook-level introduction to GPs in \citet{ras06}, and a more detailed description of the GP regression process as applied to astrophysical time-series in the appendix of \citet{gib12b}.

Under a GP model, the joint probability distribution of any finite set of $N$ outputs (in the present example, out-of-transit brightness measurements for HD\,189733) is assumed to be a multi-variate Gaussian distribution. This distribution is fully specified by an $N$-element vector of mean values, and an $N \times N$ covariance matrix. In the present application, the mean vector was assumed to be constant\footnote{In fact, the time-series was median-subtracted before modelling, and the mean vector was set to zero.} and the covariance matrix is used to describe both the intrinsic brightness variations of the star and the observational noise.

The elements of the covariance matrix are specified by a covariance function or kernel, which takes a pair of inputs (in the present case, the observation times) and returns the covariance between the corresponding pair of outputs. This function is controlled by a number of parameters, which are known as the hyper-parameters of the GP, since they specify the covariance properties of the GP, rather than directly specifying an output value at a particular time. Together, the kernel function and hyper-parameters specify a prior distribution over random functions sharing the same covariance properties. Once this prior is conditioned on the data, it gives a probability distribution for the expected output(s) at any given set of input(s), which is also Gaussian.
The choice of kernel function and of hyper-parameters are clearly critical, and need to reflect whatever information we have about the underlying physical process. As we will see below, there are many different ways of comparing kernels and hyper-parameters, ranging from a fully Bayesian to maximum likelihood and other, more ad-hoc methods.

\subsection*{Constructing a kernel for star spot variability}

We seek a kernel which appropriately describe the kind of brightness variations arising from a time-evolving distribution of spots on a rotating stellar surface.
We take the following commonly used kernels as our basic building blocks.
\emph{White noise} with standard deviation $\sigma$ is represented by:
\begin{equation}
\label{eq:WN}
k_{\rm WN}(t,t') = \sigma^2 \mathbf{I},
\end{equation}
where $\mathbf{I}$ is the identity matrix. The \emph{squared exponential} (SE) kernel is given by:
\begin{equation}
\label{eq:SE}
k_{\rm SE}(t,t') = A^2 \exp\left( - \frac{(t-t')^2}{2l^2} \right)
\end{equation}
where $A$ is an amplitude and $l$ is a length scale. This gives rather smooth variations with
a typical time-scale of $l$ and r.m.s. amplitude $A$. The \emph{rational quadratic} (RQ) kernel is given by:
\begin{equation}
\label{eq:RQ}
k_{\rm RQ}(t,t') = A^2 \left( 1 + \frac{(t-t')^2}{2\alpha l^2} \right)^{-\alpha}
\end{equation}
where $\alpha$ is known as the index. \cite{ras06} show that this is equivalent to a scale mixture of SE
kernels with different length scales, distributed according to a Beta
distribution with parameters $\alpha$ and $l^{-2}$. This gives
variations with a range of time-scales, the distribution peaking
around $l$ but extending to significantly longer periods (but remaining
rather smooth). When $\alpha \rightarrow \infty$, the RQ reduces to the
SE with length scale $l$.
There are many more types of covariance
functions in use, including some (such as the Mat{\`e}rn family) which
are better suited to model rougher, less smooth variations. However,
the SE and RQ kernels already offer a great degree of freedom with
relatively few hyper-parameters, and covariance functions based on
these are sufficient to model the data of interest satisfactorily.

A periodic covariance function can be constructed from any kernel
involving the squared distance $(t-t')^2$ by replacing the latter with
$\sin^2[\pi (t-t')/P]$, where $P$ is the period. For example, the
following:
\begin{equation}
\label{eq:perSE}
k_{\rm per,SE}(t,t') = A^2 \exp\left( - \frac{\sin^2[\pi (t-t') / P]}{2L^2} \right)
\end{equation}
gives periodic variations which closely resemble samples drawn from a
squared exponential GP within a given kernel. The length scale $L$ is
now relative to the period, and letting $L \rightarrow \infty$ gives
sinusoidal variations, whilst increasingly small values of $L$ give
periodic variations with increasingly complex harmonic content.
Similar periodic functions could be constructed from any kernel. Other
periodic functions could also be used, so long as they give rise to a
symmetric, positive definite covariance matrix -- $\sin^2$ is merely
the simplest.

As described by \cite{ras06}, valid covariance
functions can be constructed by adding or multiplying simpler
covariance functions. Thus, we can obtain a quasi-periodic kernel
simply by multiplying a periodic kernel with one of the basic kernels
described above. The latter then specifies the rate of evolution of
periodic signal. For example, we can multiply equation~\ref{eq:perSE}
with yet another SE kernel:
\begin{equation}
\label{eq:QPSE}
k_{\rm QP,SE}(t,t') = A^2 \exp\left( - \frac{\sin^2[\pi (t-t') / P]}{2L^2} - \frac{(t-t')^2}{2l^2}\right)
\end{equation}
to model a quasi-periodic signal with a single evolutionary
time-scale $l$.
In the case of activity-induced stellar brightness variations, which
are caused by the evolution and rotational modulation of active
regions, one may expect a range of both periodic covariance scales $L$
and evolutionary time-scales $l$, corresponding to different active
region sizes and life-times respectively. This can be achieved by
replacing one or both of the SE kernels in equation~\ref{eq:QPSE} by
RQ kernels.

Finally, we can also allow for correlated noise on short to moderate
time-scales by including a separate, additive SE or RQ kernel.

\subsection*{Training the GP and comparing kernels}

The marginal likelihood, which is the product of the predictive probabilities for the observed outputs, provides a `goodness of fit' measure which can be used to optimise the hyper-parameters for a given kernel (this is known as training the GP). When doing this, we exploited the fact that simple analytical expressions exist for the derivatives of the marginal likelihood with respect to the hyper-parameters to speed up the optimisation using conjugate gradient methods.

However, the marginal likelihood surface can be rather complex, specially for (quasi-)periodic kernels. This means that there is a tendency for the optimiser used to become trapped in local optima. When studying a particular dataset, as in the present case, this is circumvented by (a) carefully choosing the initial guesses for the hyper-parameters, based on a visual inspection of the data, and (b) repeating the GP training process using different initial guesses.

Ideally, we would prefer to map out the entire marginal likelihood surface, to ensure that we have found the global optimum, and to enable us to marginalise with respect to the hyper-parameters in order to compare different kernels. There are several ways of doing this: evaluating the marginal likelihood at a grid of points in the hyper-parameter space, using global optimisers, Markov Chain Monte Carlo (MCMC) methods, or Bayesian quadrature (which consists in modelling the likelihood surface itself, for example using a GP, in order to evaluate the integrals involved in the marginalisation process). Each evaluation of the marginal likelihood is computationally costly, making MCMC approaches prohibitive. We experimented with global optimisers, but with limited success. The Bayesian quadrature option is certainly the most promising, but a significant amount of code development would be needed to implement the necessary nested hierarchical models. Therefore, the results presented in this document are based on grid sampling around a manually selected initial guess, combined with a local optimiser.

We used two methods to compare different kernels: comparing the marginal likelihood obtained with the best-fit hyper-parameters in each case, and leave-one-out cross-validation (LOO-CV). Cross-validation consists in training the model on a subset of the data and then testing its ability to predict the remaining subset. In the case of GP, this is done by measuring the likelihood of the test set given the partially trained GP, as given by the predictive distribution for that subset. This process is carried out repeatedly, excluding different subsets in turn, and multiplying the results together to obtain a `pseudo-likelihood'. Rather than measuring the `goodness of fit', as the marginal likelihood does, this pseudo-likelihood measures the predictive ability of the model. Given that our task is making predictions at times where we do not have observations, this seems an appropriate way of comparing kernels. LOO-CV is a special case where each data point is `left-out' in turn. This is generally prohibitive computationally, as the model must be trained anew for each data point that is excluded. However, in the case of GPs there is a neat shortcut that allows the pseudo-likelihood for LOO-CV to be computed directly from the inverse of the full covariance matrix, making it a workable proposition.

\subsection*{Choosing a model for HD\,189733}

We modelled the HD\,189733 data with all the individual kernels detailed above, as well as a number of different combinations.  We used LOO-CV to compare different kernels, but we also performed a careful visual examination of the results, generating predictive distributions over the entire monitoring period and over individual seasons. Interestingly, we note that, in practice, LOO-CV systematically favours the simplest kernel which appears to give good results by eye.

We experimented with various combinations of square-exponential (SE) and rational-quadratic (RQ) kernels to form quasi-periodic models, and found that using an RQ kernel for the evolutionary term significantly improves the LOO-CV pseudo-likelihood relative to an SE-based evolutionary term. It makes very little difference to the mean of the predictive distribution when the observations are well-sampled, but it seems to increase the predictive power of the GP away from observations, as it allows for a small amount of covariance on long time-scales, even if the dominant evolution time-scale is relatively short. On the other hand, we were not able to distinguish between SE and RQ kernels for the periodic term (the two give virtually identical best-fit marginal likelihoods and pseudo-likelihoods), and therefore opted for the simpler SE kernel.

To describe the observational noise, we experimented with a separate, additive SE kernel as well as a white noise term. However, we found that this did not significantly improve the marginal or pseudo-likelihood, and the best-fit length-scale was comparable to the typical interval between consecutive data points. We therefore reverted to a white-noise term only. The final kernel was thus:
\begin{equation}
\label{eq:QPGP_final}
\begin{array}{ll}
k_{\rm QP,mixed}(t,t') = & A^2 \exp\left( - \frac{\sin^2[\pi (t-t') / P]}{2L^2}  \right) \\
& \times \left( 1 + \frac{(t-t')^2}{2\alpha l^2} \right)^{-\alpha} + \sigma^2 \mathbf{I}.
\end{array}
\end{equation}

The best-fit hyperparameters were $A=6.68$\,mmag, $P=11.86$\,days, $L=0.91$, $\alpha=0.23$, $l=17.80$\,days, and $\sigma=2.1$\,mmag. Our period estimate is in excellent agreement with \cite{hen08}. We note that very similar best-fit hyper-parameters were obtained with the other kernels we tried (where those kernel shared equivalent hyper-parameters). The relatively long periodic length-scale $L$ indicates that the variations are dominated by fairly large active regions. The evolutionary term has a relatively short time-scale $l$ (about 1.5 times the period) but a relatively shallow index $\alpha$, which is consistent with the notion that individual active regions evolve relatively fast, but that there are preferentially active longitudes where active regions tend to be located (as inferred from better-sampled long-duration CoRoT light curves for similarly active stars).

\subsection*{Results}

Once the GP has been trained and conditioned on the available data, we can compute a predictive distribution for any desired set of input times. This predictive distribution is a multi-variate Gaussian, specified by a mean vector and a covariance matrix \citep[see e.g.][for details]{gib12b}. Here we are interested in estimating the difference between the predicted flux at two input times, which is just the difference between the corresponding elements of the mean vector. We also want an estimate of the uncertainty associated with this predicted difference. This can be obtained directly from the covariance matrix of the predictive distribution:
\begin{equation}
\label{eq:UncDiff}
\sigma^2_{y_2|y_1} = {\rm cov}(x_1, x_1) + {\rm cov}(x_2, x_2) - 2 {\rm cov}(x_1, x_2).
\end{equation}

The results are reported in Table~\ref{dimming}. We take as our reference the maximum flux predicted by the GP at any time throughout the $5+$ years of monitoring with the APT. Assuming that the spots are dark, this represents a lower limit on the spot-free flux. The flux differences relative to this level can readily be converted to differences relative to another, arbitrarily higher flux by adding a constant to all of them. Note that we worked in magnitude units when training and conditioning the GP, but have been careful to convert the results to flux units before tabulating them. We computed the flux drop values assuming a spot temperature of 4000\,K. Changing the spot temperature changes the training data very slightly, because it changes the conversion factor between the Wise and APT bandpasses. However, this makes only a very small difference to the flux estimates for transit observations occurring during the same season as the Wise observations, and none at all for the other transit observations, so we have not tabulated the estimated flux differences assuming other spot temperatures. This does not preclude trying out various spot temperatures when converting the flux drops reported in  Table~\ref{dimming} to the bandpasses of the various transit observations.

\section[]{Uncertainties on phase curve amplitudes due to stellar activity}

In \citet{knu12}, the quasi-periodic GP trained on the ground-based APT data (as described in Appendix~A) was used to predict whether the stellar flux during the Spitzer observations was linear or underwent an inflection, and if so what direction the inflection went. An MCMC code was then used to adjust the coefficients of a model that contained these terms for the stellar variability, plus a series of systematics terms, plus planetary phase-curve terms. The final uncertainties on the phase curve amplitudes were derived directly from these MCMC runs.

Here we try to address two questions:
\begin{enumerate}
\item Is the predictive capability of the GP model sufficient to use it in that way?
\item To what extent can HD\,189733's variability mimic a phase curve? How does that compare with the
errors on the amplitude reported in the K12 paper?
\end{enumerate}

\subsection*{Predictive power of the GP}

\begin{figure}
\resizebox{\linewidth}{!}{\includegraphics{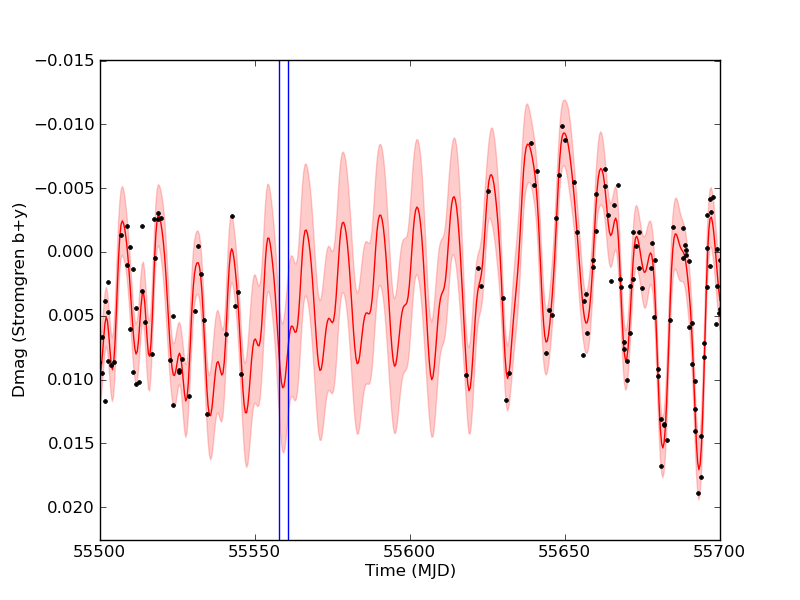}
\includegraphics{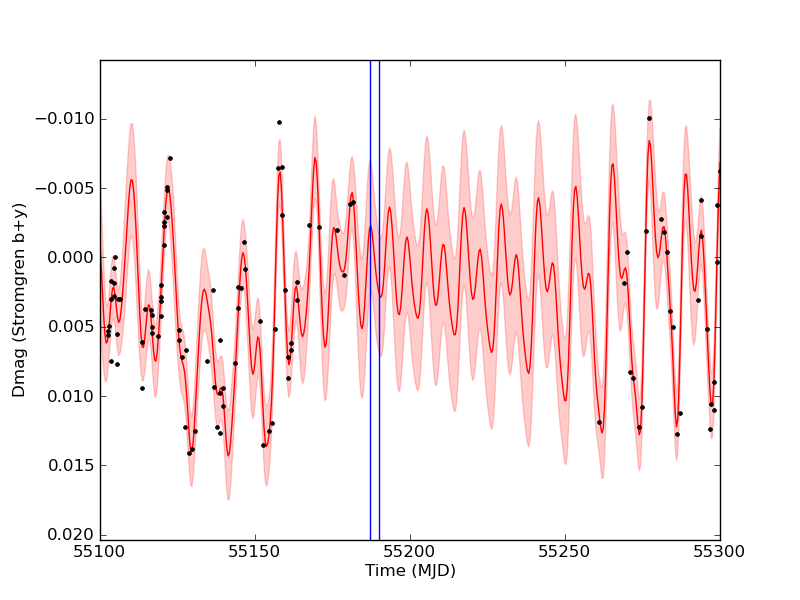}}
\caption{Timing of the \Spitzer\ phase curve observations at 3.6 (left) and 4.5\,\micron\ (right), indicated by the vertical blue lines, compared to the APT observations (black points). The mean and 1-$\sigma$ confidence interval of the predictive distribution of the GP model as shown by the red line and pink shaded area, respectively. Reproduced from K12.}
\label{timingGP}
\end{figure}

As shown in Fig.~\ref{timingGP}, reproduced from K12, the \Spitzer\ observations at 3.6 and 4.5\,\micron\ each occur after the end of a ground-based monitoring season.  As a result, the GP's ability to predict the overall level, amplitude and phase of the quasi-periodic modulation of the stellar flux during the \Spitzer\ observations is somewhat limited. This is illustrated by Fig.~\ref{randomGP}, where we have drawn random samples from the GP around the time of the \Spitzer\ observations.

\begin{figure}
\resizebox{\linewidth}{!}{\includegraphics{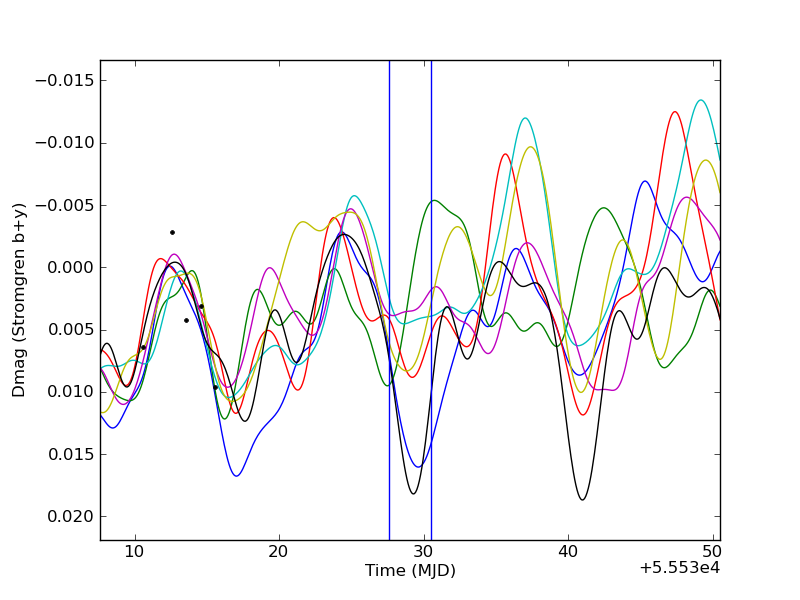}\includegraphics{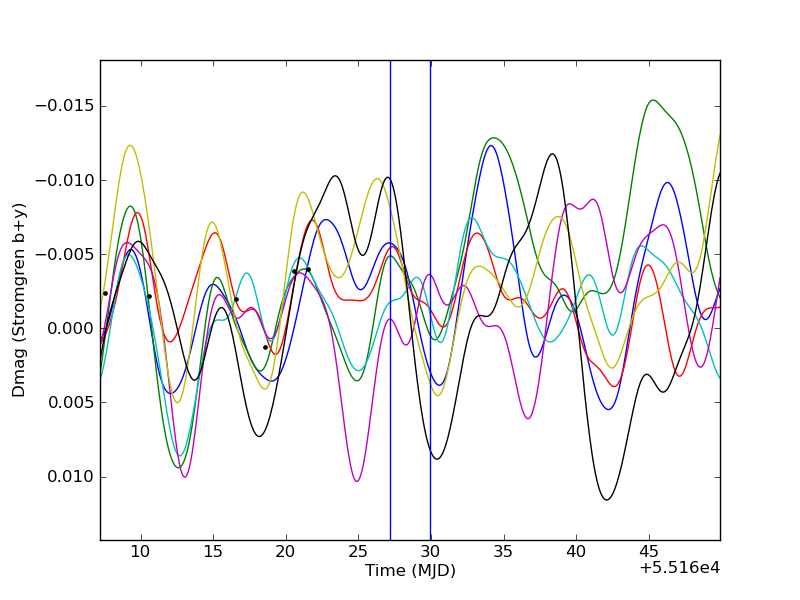}}
\caption{Random realisations of the GP (coloured lines) drawn around the time of the \Spitzer\ observations.}
\label{randomGP}
\end{figure}

Nonetheless, this exercise can be used to check what fraction of the time the GP samples are better fit by a linear or quadratic function, and to check if the quadratic term tends to be positive or negative. This is what was done in K12 in the case of the 3.6-\micron\ passband, the GP samples tend to undergo a minimum, so K12 used a quadratic term and forced it to be positive. In the case of the 4.5-\micron\ passband, the GP samples tend to decrease monotonically, so K12 used only a linear term.

One limitation of the approach above is that the samples are still drawn from a single GP, whose covariance hyper-parameters were trained on the APT data. The uncertainty on these hyper-parameters is not accounted for. In particular, the rather sparse sampling of the APT data (one point every day or two during the observing seasons) means that we cannot constrain the periodic length scale well. The best-fit GP has a fairly large periodic length scale, making it rather smooth, with only one or two inflections per period. But if there were more short-term variations in a period, we would not be able to distinguish them from white noise using the APT data alone. Since the phase curve signal operates on the time scale of the planet orbital period (2.2 days), these short-term variations would affect the phase curve measurements.

We can use the 2007 observations of HD\,189733 by the MOST space mission (kindly provided by Bryce Croll), which have much better precision and time sampling than the APT data, and were taken outside the APT observing season, to test the importance of the limitations noted above. The MOST data provides continuous coverage of the brightness of HD\,189733 over one month. The transits of HD\,189733b had already been removed from these data, and a linear trend of order $10^{-4}$ mag per day, suspected to be of instrumental origin, had been subtracted. We scaled the MOST magnitude variations by a factor 1/0.987 to account for the difference between the MOST and APT passbands (the former was crudely approximated by a top-hat throughput function from 350 to 700\,nm), and added a constant to the MOST dataset to bring its median magnitude level with that of the surrounding two semesters of APT observations. Finally, we binned up the MOST data (from  $\sim$1 point per hour to $\sim$2 points per day, resulting in a total of 69 data points).

\begin{figure}
\resizebox{\linewidth}{!}{\includegraphics{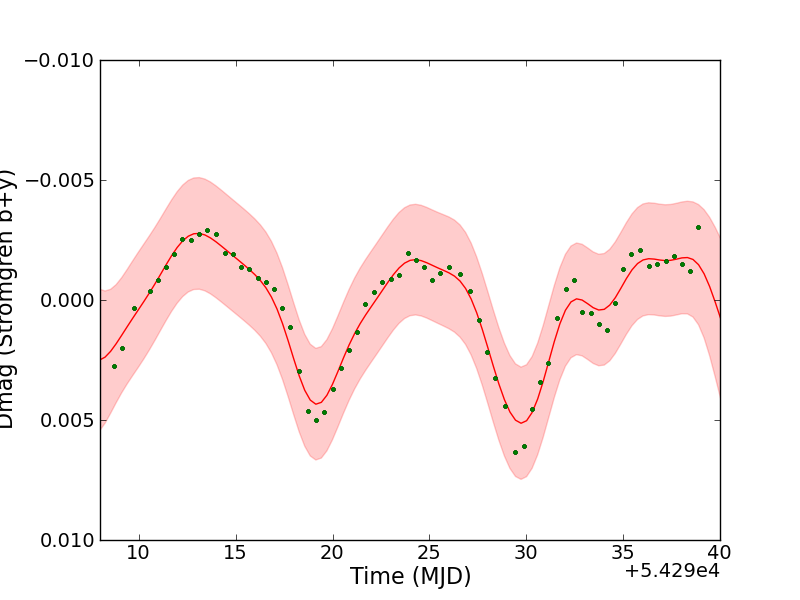}}
\caption{MOST 2007 observations of HD\,189733 (green dots) together with the predictive mean (red line) and standard deviation (pink shaded area) of the GP model trained on the APT data alone, but conditioned on the MOST data. Note that the standard deviation includes the white noise level of the APT data, which is significantly larger than that of the MOST data.}
\label{MOST}
\end{figure}

In Fig.~\ref{MOST}, we have trained the GP (adjusted its hyper-parameters) on the APT data, but then conditioned it (adjusted the mean and covariance, but keeping the hyper-parameters fixed) on the MOST data (shown in green, after removing the transits, and binning to two points per day for clarity). The GP is clearly unable to follow the more rapid variations in the MOST flux. This shows that the GP trained on the APT data does not  capture the short-timescale variability. In fact the variability it fails to capture is precisely on the 1-2 day timescales relevant to the phase curve measurements. Additionally, we also performed a more stringent test, where the GP was trained and conditioned only on the APT data, and compared to the MOST observations without conditioning it on the latter. The results are shown in Fig.~\ref{MOST2}. They show very clearly that, although the gap between the APT and MOST observations is only around one stellar rotation period (much like the gap between the APT and \Spitzer\ observations), this is enough for the GP to lose its ability to predict the phase and, to a lesser extent, the amplitude of the signal. In the light of this new test, we are compelled to question the validity of using the GP to constrain the \Spitzer\ fits, even in the very conservative fashion adopted in K12.

\begin{figure}
\resizebox{8cm}{!}{\includegraphics{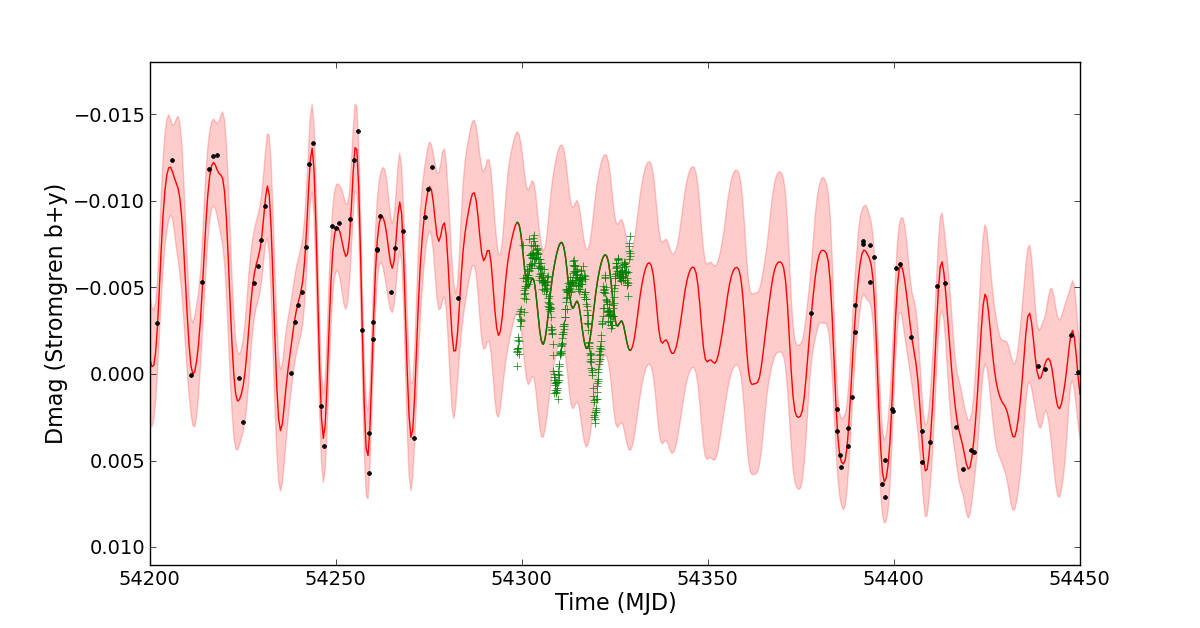}}
\caption{MOST (green dots) and APT (black dots) observations of HD\,189733 (green dots) together with the predictive mean (red line) and standard deviation (pink shaded area) of the GP model trained and conditioned on the APT data alone.}
\label{MOST2}
\end{figure}

\subsection*{Estimating the activity-induced error on the phase curve amplitude:}

We know that stellar variability contributes to the uncertainty on the phase-curve amplitude, but we have shown, using the MOST data, that the GP cannot be used to \emph{correct} for this effect reliably. On the other hand, the same MOST observations can be used to \emph{evaluate} the contribution of activity to the uncertainty on the phase curve amplitude, again making use of the MOST data. Stellar variability is expected to be by far the dominant contribution to the brightness fluctuations observed with MOST; any orbital phase variations of planetary origin (arising, from example, from star light reflected by the planet) are expected to be tiny, and certainly much smaller than the infrared phase-curve amplitudes. Thus, if all we observe is stellar variability, but we try to model is as a combination of a planetary and a stellar signal, as done in K12, what phase curve amplitudes do we obtain? This can be used to estimate the contribution from stellar variability to the error on the phase curve amplitude (after applying a suitable scaling for the difference in bandpasses).

\begin{figure}
\resizebox{\linewidth}{!}{\includegraphics{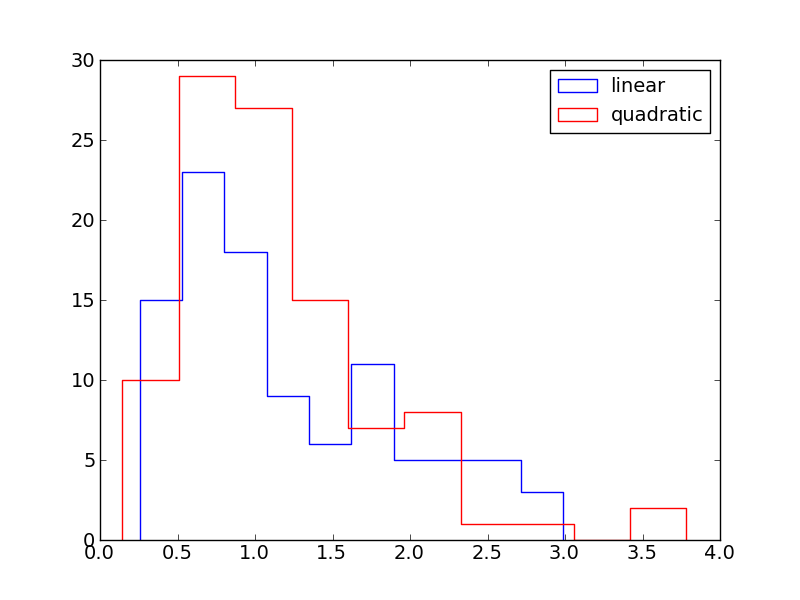}}
\caption{Histograms of the amplitude of the phase-curve component of the fitted model (in parts per thousand). The two colours correspond to linear and quadratic models for the variability terms. The x-axis is in parts per thousand.}
\label{histpc}
\end{figure}

To perform this test, we took the (un-binned) MOST time-series, extracted sections lasting 2.7 days (approximate duration of the \Spitzer\ phase curve observations) and fit them with a model containing a polynomial (order 1 or 2) to represent the stellar variability, and a sum of sines and cosines at the period and half the period of the planet, to represent a potential planetary phase curve. The phase of the sinusoidal terms was always fixed so that the transit would have occurred in the middle of the data segment used (as in the \Spitzer\ observations). Figure~\ref{histpc} shows histograms of the amplitude of the phase-curve component of the fitted model. Regardless of the order of the polynomial term, the stellar variability can mimic phase curve signals at the level of at least 0.1\% in the optical. This needs to be divided by a factor of 3.7 and 3.9 to convert it from the MOST to the Spitzer channel 1 and 2 bandpasses respectively. To summarise, stellar variability induces an uncertainty on the \Spitzer\ phase curve amplitudes of at least 0.027 and 0.025\% at 3.6 and 4.5~\micron, respectively. For comparison, the phase curve amplitude and uncertainties adopted in K12, derived from constrained MCMC fits to the \Spitzer\ light curves, were $0.124 \pm 0.006$ and $0.098 \pm 0.009$\% respectively.

To understand the implications for this in terms of atmospheric physics, we need to convert the phase curve amplitudes to temperature contrasts. This gives uncertainties of 150 and 78\,K for the brightness temperatures at 3.6 and 4.5 \micron, respectively. Thus the new value for the day-side brightness temperatures measured during the secondary eclipses become: $1328 \pm 150$\,K at 3.6 \micron, and $1192 \pm 78$\,K at 4.5 \micron. Our revised error estimates do not bring the detection of a phase modulation into question (both remain $>4\sigma$), but it does substantially decrease the significance level of the difference between the day-night contrast in the two passbands.

\section{Slope of the transmission spectrum with settling grains}

We calculate the slope of the transmission spectrum, in other words the dependence of the effective transit radius $r$ as a function of wavelength $\lambda$, when the opacity is dominated by Rayleigh scattering from grains of condensates of effective size $a$, with the density and maximum size of the grains varying with pressure (``settling grains'' case in Section~\ref{settling}). 

We use the usual approximation that the effective transit radius corresponds to the height in the atmosphere at which the opacity reaches a critical value $\tau_{\rm crit}$.

If $\xi(z)$ is the abundance of grains at height $z$ in the atmosphere, then

\begin{equation}
\tau(z,\lambda) \propto \int_a n(z) \sigma(a,\lambda) {\rm d} a= \int_a \rho(z) \xi(z) \sigma(a,\lambda) {\rm d}  a
\end{equation}
where $n$ is the number density of grains, $\rho$ the atmospheric density and $\sigma$ the cross-section of individual grains.

For Rayleigh scattering, 
\begin{equation}
\sigma(a,\lambda) \propto  a^6 \lambda^{-4} 
\end{equation}

Because of the very steep dependence of the cross-section on grain size, the largest grains at a given height dominate. We therefore use:
\begin{equation}
\sigma(z,\lambda) \propto  a_{\rm max}^6 (z) \lambda^4 
\end{equation}
thus

\begin{equation}
\tau_{\rm crit} \propto \rho(z) \xi(z)  a_{\rm max}^6 (z) \lambda^{-4}
\end{equation}

In the ideal-gas, isothermal approximation, 
\begin{equation}
\rho(z) \propto p(z)
\end{equation}
 and
 \begin{equation}
 p(z) \propto e^{-z/H}
 \end{equation}
 where $p$ is the gas pressure and $H$ is the atmospheric scale height.
 
For the scenarios in Section~\ref{go} we assume $a_{\rm max}(z) \propto p(z)$ (settling grains in the molecular regime), and use the following dependence of $a_{\rm max}$ on $p$:
\begin{equation}
\xi(a_{\rm max}) \propto a^\beta
\end{equation}
with $\beta = 0$ corresponding to a flat distribution of grain sizes, and $\beta = -3$ corresponding to an equal partition of mass across all grain sizes. 
This gives
\begin{eqnarray*}
\tau_{\rm crit} &=& \rho(z) \xi(z)  a_{\rm max}^6 (z) \lambda^{-4} \\
&=& p(z) p(z)^\beta p(z)^6 \lambda^{-4}  = p^{(7+\beta)} \lambda^{-4} \\
&=& e^{-(7+\beta)z/H} \lambda^{-4}
\end{eqnarray*}

Taking the logarithm gives:

\begin{equation}
z= -\frac{4}{7+\beta}  H \ln \lambda
\end{equation}

Therefore the slope of the transmission spectrum, in units of atmospheric scale height $H$ per $\ln \lambda$, is $-4/7$ in the case of flat grain size distribution ($\beta=0$), and $-1$ in the case of constant mass fraction distribution of grain sizes ($\beta=-3$).

\label{lastpage}
\bsp

\end{document}